\documentclass[aps,prd,reprint,nofootinbib,longbibliography,superscriptaddress]{revtex4-2}

\usepackage[T1]{fontenc}
\usepackage[utf8]{inputenc}
\usepackage{lmodern}
\usepackage{amsmath,amssymb,amsthm,mathtools,bm}
\usepackage{physics}
\usepackage{microtype}
\usepackage{hyperref}
\usepackage{xcolor}

\hypersetup{colorlinks=true,citecolor=blue,linkcolor=blue,urlcolor=blue}

\newcommand{\Th}{\Theta}
\newcommand{\D}{\mathrm{D}}
\newcommand{\FF}{\mathcal{F}}

\begin{document}

\title{First-order thermodynamics of multi-scalar-tensor gravity}

\author{David S. Pereira}
\email{djpereira@ciencias.ulisboa.pt}
	\affiliation{%
		Departamento de F\'{i}sica, Faculdade de Ci\^{e}ncias da Universidade de Lisboa, Campo Grande, Edif\'{\i}cio C8, P-1749-016 Lisbon, Portugal}
	\affiliation{Instituto de Astrof\'{\i}sica e Ci\^{e}ncias do Espa\c{c}o, Faculdade de
		Ci\^{e}ncias da Universidade de Lisboa, Campo Grande, Edif\'{\i}cio C8,
		P-1749-016 Lisbon, Portugal;\\
	}%

\date{\today}

\begin{abstract}
We formulate a first-order thermodynamic description of Jordan-frame tensor--multi-scalar gravity. From the Einstein-like field equations we obtain the exact covariant $1+3$ decomposition of the geometric sector and write it as an effective imperfect fluid. The thermodynamic interpretation is understood in the first-order Eckart sense: the effective temperature, conductivity, entropy current, and entropy production are meaningful only on branches where the geometric dissipative variables satisfy the appropriate matching and integrability conditions. In a generic frame, the heat flux admits the exact decomposition $q_a^{(g)}=-\chi(a_a+W_a)$, with $\chi=-\dot{\FF}/(8\pi\FF)$ and with $W_a$ encoding the residual temperature-gradient sector. In the $\FF$-comoving frame this gives the inertial variable $\chi_{\FF}\equiv K_{\FF}T_{\FF}$ together with a generally nonvanishing spatial contribution $W_a^{(\FF)}$, sourced by scalar directions not aligned with the coupling. Thus the multi-field thermal description is not generically reducible to a single $KT$-type quantity. We derive transport equations for $\chi_{\FF}$, for the field-space thermal vector $\chi^A$ and covector $\chi_A$, and for the residual gradient sector. We also introduce the diagnostics $\mathfrak D_\chi=\chi_A\chi^A$ and $\mathfrak D_{\rm grad}=\mathcal B_{AB}\D_a^{(\FF)}\phi^A\D^a_{(\FF)}\phi^B$. Their interpretation depends on the kinetic matrix $\mathcal B_{AB}$: they are canonical contractions when $\mathcal B_{AB}$ is nondegenerate, but become nonnegative norm-like diagnostics only when $\mathcal B_{AB}$ is positive definite; if $\mathcal B_{AB}$ is degenerate, no model-independent inverse field-space metric or full scalar norm is available without extra structure. With this qualification, these diagnostics show that freezing the effective coupling is, in general, weaker than full relaxation to the GR sector. Finally, we construct the entropy current and entropy production in the coupling frame, state the assumptions required for nonnegative entropy production, and show that homogeneous cosmology suppresses the spatial sector while retaining nontrivial time-like multi-scalar thermal dynamics.
\end{abstract}
\maketitle
\section{Introduction}\label{sec:intro}
General Relativity (GR) remains extraordinarily successful across a wide range of scales, from Solar-System experiments and binary pulsars to the strong-field and gravitational-wave regime~\cite{Will:2014kxa,Berti:2015itd}. Nevertheless, there are compelling reasons to investigate consistent extensions of Einstein gravity. On the observational side, the origin of cosmic acceleration and the broader dark-sector problem continue to motivate departures from the minimal GR picture, while on the theoretical side one would like to understand how rigid the Einstein description really is once additional gravitational degrees of freedom are allowed~\cite{Clifton:2011jh,CANTATA:2021asi,CosmoVerseNetwork:2025alb}. Scalar-tensor theories occupy a central place in this program because they provide one of the simplest and most systematic extensions to GR while preserving a transparent covariant field-theoretic structure~\cite{Brans:1961sx,Horndeski:1974wa,Faraoni:2004pi,Clifton:2011jh}. They also serve as a common language in which many modified-gravity models and effective descriptions of extra gravitational sectors can be cast.

A natural step beyond the usual one-field setting is to allow several scalar degrees of freedom. In that case the scalar sector is no longer described by a single extra variable, but by a nontrivial field space endowed with its own metric and couplings. Tensor--multi-scalar gravity was systematized in this geometric form by Damour and Esposito-Far\`ese~\cite{Damour:1992we}, and early studies already showed that multi-field scalar-tensor models can display cosmological and gravitational phenomena absent in the one-field case~\cite{Berkin:1993bt}. Such theories are also well motivated from high-energy constructions, since dimensional reduction and string compactifications generically produce several moduli and dilaton-like fields in the four-dimensional effective description~\cite{Grana:2005jc}. Multi-field scalar-tensor theories have been investigated across a broad range of physical settings, and the results consistently point to phenomena with no strict one-field counterpart. In early-universe cosmology they have been used to describe curved trajectories in field space, genuinely multifield background evolution, entropy/isocurvature perturbations, adiabatic--entropy transfer, non-Gaussian signatures, and reheating dynamics~\cite{Starobinsky:2001xq,Wands:2007bd,Byrnes:2006fr,Kaiser:2012ak,Kaiser:2013sna,Karamitsos:2017elm,DeCross:2015uza,DeCross:2016fdz,DeCross:2016xpj,Nguyen:2019kbm}. More recent work has highlighted robust many-field attractor behavior in nonminimally coupled inflation, the extraction of cosmological observables in multifield settings, and the extension of screening mechanisms to genuinely multi-field dilaton scenarios~\cite{Christodoulidis:2025wew,Bolshakova:2025xay,Brax:2026ttj}. At late times, multifield scalar-tensor models have been studied as dark-energy and dark-sector frameworks, where they can sustain accelerated expansion, modify clustering and structure growth, and be confronted directly with CMB, BAO, supernova, and cosmic-chronometer data~\cite{Akrami:2020zlw,Smith:2024mbp,Smith:2024mna,Smith:2025wxd,Gallego:2026xqq}. Beyond cosmology, they have also been explored in the weak-field regime, in exact cosmological solutions and phase-space analyses, in junction conditions and matched geometries, in wormhole and compact-star configurations, in post-Newtonian dynamics, and in gravitational-wave propagation with additional scalar channels~\cite{Rosa:2021lhc,Rosa:2017jld,Rosa:2019ejh,Rosa:2021mln,Rosa:2018jwp,Aliannejadi:2024byu,Hohmann:2016yfd,Schon:2021uhz,Gomes:2025ers,Pereira:2025rsr}. At the level of covariant gravitational theory, their effective-fluid-mixture structure has also been clarified explicitly~\cite{Miranda:2022uyk}. Taken together, these results show that tensor--multi-scalar gravity is not merely a formal extension of the one-field case, but a phenomenologically rich framework in which multiple scalar directions can affect cosmology, astrophysics, and gravitational phenomenology in qualitatively new ways.

Furthermore, over the last few years, a new perspective on scalar-tensor gravity has emerged through its connection with first-order relativistic thermodynamics~\cite{Faraoni:2018qdr,Faraoni:2021lfc,Faraoni:2021jri,Giardino:2022sdv,Faraoni:2022gry,Faraoni:2023hwu,Giardino:2023ygc,Miranda:2024dhw,Houle:2024sxs,Faraoni:2025alq,Faraoni:2025ufi,Giusti:2021sku,Faraoni:2025dex,Faraoni:2025fjq,Bhattacharyya:2025tgp,Jarv:2026dbb}. In this framework, the scalar sector is rewritten as an effective imperfect fluid and its geometric heat flux, anisotropic stress, and bulk sector are compared with Eckart-type constitutive relations~\cite{Eckart:1940te}. This makes it possible to introduce transport coefficients, an effective temperature, an entropy current, and a notion of thermal evolution relative to the Einstein limit. Within this viewpoint, GR appears as a zero-temperature equilibrium state, while scalar-tensor configurations are interpreted as nonequilibrium excitations of the gravitational medium~\cite{Faraoni:2021lfc,Faraoni:2021jri,FaraoniGiustiEtAl2022Peculiar,Giardino:2022sdv,Faraoni:2025alq}. This thermodynamic picture has already been developed for ordinary scalar-tensor gravity, minimally coupled scalars, Horndeski cosmology, anisotropic cosmologies, and the recent thermal reinterpretation of attractor-to-GR behavior~\cite{Giardino:2022sdv,Giusti:2021sku,Faraoni:2022gry,Faraoni:2023hwu,Miranda:2024dhw,Houle:2024sxs,Giardino:2023ygc,Faraoni:2025alq}. Even though first-order relativistic thermodynamics is known to have well-known limitations concerning causality and stability~\cite{Hiscock:1985zz,Maartens:1996vi}, it still provides a sharp constitutive framework with which to diagnose departures from GR.

Moreover, the first-order character of this construction should also be kept explicit. The present paper should not be read as providing a causal, second-order thermodynamics of generic tensor--multi-scalar gravity. Rather, it gives the first-order constitutive and transport framework that must be understood before a causal extension can be attempted. Recent work has begun to extend the scalar-tensor thermal analogy beyond Eckart by using Israel--Stewart-type ideas in the ordinary one-field scalar-tensor case~\cite{Banerjee:2026ulx}. To our knowledge, however, a fully developed second-order causal formulation covering both the general Jordan-frame one-field theory and the generic multi-scalar system considered here is not yet available. The present work is therefore deliberately first-order: its purpose is to determine when the geometric multi-scalar effective fluid admits through matching an Eckart-type thermal interpretation and what additional multi-field thermal channels appear at that level.

Extending the thermodynamic framework from the one-field case to the multi-field setting is therefore important for both theoretical and practical reasons. In the one-field case, the thermodynamic description does more than recast the field equations: it provides a unified language in which departures from GR can be interpreted in terms of effective nonequilibrium variables, transport properties, and relaxation processes~\cite{Faraoni:2025alq}. In multi-field theories this becomes more relevant, because the scalar sector has a genuinely richer field-space structure and different scalar directions can play distinct dynamical roles. A multi-field thermodynamic framework can thus clarify how this enlarged gravitational sector is organized and how the approach to, or departure from, a GR-like regime should be understood. In this regard, the work of Miranda, Graham, and Faraoni is especially relevant~\cite{Miranda:2022uyk}. There it was shown that tensor--multi-scalar gravity admits an effective-fluid-mixture interpretation and that several observer congruences can be physically meaningful, including one adapted to the effective coupling. This provides precisely the structural background needed for a genuine thermodynamic analysis. What is still missing is a constitutive and transport-level thermodynamic formulation for generic Jordan-frame tensor--multi-scalar gravity, valid beyond the effectively one-field aligned sectors.

The aim of the present paper is to provide that extension. We work with a general Jordan-frame tensor--multi-scalar theory, keeping the common coupling function $\FF(\phi^A)$ and the field-space metric $\mathcal{B}_{AB}(\phi^C)$ explicit throughout. This general formulation is useful for two related reasons. First, it makes clear which aspects of the thermodynamic description are universal at the multi-field level and which belong only to restricted one-field-like sectors. Second, it allows the formalism to be applied to a generic class of theories.

Our main results can be summarized as follows. First, we obtain the exact $1+3$ decomposition of the Jordan-frame geometric sector and show that, in the coupling frame, the heat flux isolates the usual inertial quantity $\chi_{\FF}\equiv K_{\FF}T_{\FF}$ but generically retains an additional spatial piece $W_a^{(\FF)}$, which measures the residual temperature-gradient sector sourced by scalar directions not aligned with the coupling. This is a first-order constitutive statement: the existence of an effective temperature requires the corresponding Eckart matching and the local integrability condition for $W_a^{(\FF)}$. Second, we derive the transport system for $\chi_{\FF}$, for the field-space thermal vector $\chi^A$ and covector $\chi_A$, and for the gradient sector itself, showing explicitly that coupling relaxation and full multi-scalar relaxation are distinct. Third, we introduce the scalar diagnostics $\mathfrak D_\chi$ and $\mathfrak D_{\rm grad}$, which quantify the time-like and projected spatial multi-scalar sectors only under clearly stated assumptions on $\mathcal B_{AB}$. If $\mathcal B_{AB}$ is nondegenerate, these are canonical scalar contractions; if $\mathcal B_{AB}$ is positive definite, they become nonnegative norm-like diagnostics. Under this stronger assumption they lead to a precise GR-attractor criterion: freezing the effective coupling is, in general, weaker than full relaxation to the GR sector. Finally, we construct the entropy current and entropy production in the coupling frame. We explicitly distinguish the entropy-production functional obtained by first-order matching from a guaranteed second-law statement: nonnegative entropy production requires additional sign assumptions on the effective transport coefficients, or a restriction to branches where the problematic sectors vanish. We then show that homogeneous cosmology suppresses the spatial sector by symmetry while retaining nontrivial time-like multi-scalar thermal dynamics.

The paper is organized as follows. Section~\ref{sec:theory} introduces the general Jordan-frame tensor--multi-scalar theory. Section~\ref{sec:covariant} presents the exact covariant $1+3$ decomposition of the geometric source. Section~\ref{sec:thermal} studies the generic-frame constitutive matching to Eckart thermodynamics, while Sec.~\ref{sec:frames} specializes the discussion to the natural scalar-comoving frames and the aligned-gradient sector. Section~\ref{sec:transport} develops the transport analysis. Section~\ref{sec:entropy} discusses the entropy current and entropy production. Section~\ref{sec:cosmoKTF} applies the formalism to homogeneous and isotropic cosmology, and Sec.~\ref{sec:discussion} summarizes the conclusions. Throughout, we use the metric signature $(-,+,+,+)$ and units with $c=1$.

\section{Jordan-frame multi-scalar-tensor gravity}
\label{sec:theory}

We consider a general Jordan-frame multi-scalar-tensor theory (also referred to in the literature as tensor-multi-scalar gravity) described by the action~\cite{Berkin:1993bt,Hohmann:2016yfd,Miranda:2022uyk}
\begin{align}
S&=\frac{1}{16\pi}\int d^4x\sqrt{-g}\,\left[\FF(\phi^C)R-\mathcal{B}_{AB}(\phi^C)\nabla_a\phi^A\nabla^a\phi^B \right. \nonumber\\
&\left.-U(\phi^C)\right]+S_m[g_{ab},\Psi_m],
\label{eq:multiaction}
\end{align}
where capital Latin indices label the scalar fields, $\FF(\phi^C)$ is the common nonminimal coupling function, $\mathcal{B}_{AB}(\phi^C)$ is a symmetric field-space kinetic matrix, and $U(\phi^C)$ is the scalar potential. Matter is minimally coupled to the spacetime metric $g_{ab}$.

We impose
\begin{equation}
\FF>0,
\label{eq:Fpositive_multi}
\end{equation}
so that the effective gravitational coupling remains positive. The kinetic matrix $\mathcal B_{AB}$ plays several logically distinct roles in what follows, and it is useful to separate them from the outset. The lower-index scalar equations obtained directly from the action are meaningful for any symmetric $\mathcal B_{AB}$, including degenerate cases. If $\mathcal B_{AB}$ is invertible, one may introduce $\mathcal B^{AB}$ and rewrite the scalar equations in field-space covariant form. If, in addition, $\mathcal B_{AB}$ is positive definite on the physical scalar sector, then it defines a genuine Riemannian field-space metric and contractions such as $\mathcal B_{AB}X^A X^B$ are nonnegative norm-like quantities. We shall keep these assumptions distinct below. In particular, all formulas involving $\mathcal B^{AB}$, $\Gamma^{A}{}_{BC}(\mathcal B)$, $\FF^{,A}$, or $U^{,A}$ should be read as restricted to the nondegenerate branch, while statements interpreting $\mathcal B_{AB}$-contractions as positive magnitudes require the stronger positive-definiteness assumption.

Varying Eq.~\eqref{eq:multiaction} with respect to the metric gives
\begin{align}
\FF G_{ab}&=8\pi T^{(m)}_{ab}+\nabla_a\nabla_b\FF-g_{ab}\Box\FF-\frac12 U\,g_{ab}\nonumber\\
&+\mathcal{B}_{AB}\left(\nabla_a\phi^A\nabla_b\phi^B-\frac12 g_{ab}\nabla_c\phi^A\nabla^c\phi^B\right).
\label{eq:metricmulti}
\end{align}
where, due to the Bianchi identities, the matter stress tensor defined as $T_{ab}\equiv-\frac{2}{\sqrt{-g}}\frac{\delta S_m}{\delta g^{ab}}$ obeys
\begin{equation}
\nabla^b T^{(m)}_{ab}=0,
\label{eq:mattercons_multi}
\end{equation}
because matter is minimally coupled to $g_{ab}$.

Variation with respect to the scalar fields $\phi^A$ gives
\begin{align}
2\mathcal{B}_{AB}\Box\phi^B
&=-
\left(
2\mathcal{B}_{AB,C}-\mathcal{B}_{BC,A}
\right)
\nabla_a\phi^B\nabla^a\phi^C \nonumber\\
&-\FF_{,A}R+U_{,A},
\label{eq:scalarmulti}
\end{align}
where a comma denotes partial differentiation with respect to the scalar fields, e.g.\
$\mathcal{B}_{AB,C}\equiv \partial \mathcal{B}_{AB}/\partial\phi^C$.
The first term is the ordinary d'Alembertian contribution, the second arises because the kinetic matrix
$\mathcal{B}_{AB}(\phi)$ depends on the fields themselves, and the last two terms encode, respectively, the coupling of the scalar sector to the Ricci scalar through $\FF(\phi)$ and the force generated by the potential $U(\phi)$.

When $\mathcal{B}_{AB}$ is invertible, it can be regarded as a metric on the scalar-field manifold. In that case Eq.~\eqref{eq:scalarmulti} can be rewritten in the field-space covariant form
\begin{equation}
\Box\phi^A
+\Gamma^{A}{}_{BC}(\mathcal{B})\nabla_a\phi^B\nabla^a\phi^C
+\frac12\,\mathcal{B}^{AB}\left(\FF_{,B}R-U_{,B}\right)=0,
\label{eq:scalarmulticov}
\end{equation}
where
\begin{equation}
\Gamma^{A}{}_{BC}(\mathcal{B})
\equiv
\frac12\mathcal{B}^{AD}
\left(
\mathcal{B}_{DB,C}
+\mathcal{B}_{DC,B}
-\mathcal{B}_{BC,D}
\right)
\label{eq:fieldspaceGamma}
\end{equation}
is the Levi-Civita connection associated with the field-space metric $\mathcal{B}_{AB}$. This rewriting is not available in a model-independent way if $\mathcal{B}_{AB}$ is degenerate. In that case Eq.~\eqref{eq:scalarmulti} remains the fundamental scalar equation. Degenerate directions may correspond to constraints, nondynamical combinations, or other model-dependent structures, and one cannot define a unique inverse field-space metric without supplying additional information. Thus the thermodynamic analysis below can still be formulated at the level of the effective stress tensor and lower-index constitutive quantities, but the field-space covariant scalar equation and norm-like diagnostics require the appropriate nondegeneracy or positivity assumptions.

Equation~\eqref{eq:scalarmulticov} makes the geometric structure of the scalar dynamics manifest. The combination
\begin{equation}
\Box\phi^A+\Gamma^{A}{}_{BC}\nabla_a\phi^B\nabla^a\phi^C,
\end{equation}
is the field-space covariant generalization of the Klein-Gordon operator, showing that the scalar fields evolve as a coupled system on a curved target space. The remaining term,
\[
\frac12\,\mathcal{B}^{AB}\left(\FF_{,B}R-U_{,B}\right),
\]
acts as an effective force, sourced jointly by the nonminimal coupling function $\FF(\phi)$ and by the scalar potential $U(\phi)$.

Taking the trace of Eq.~\eqref{eq:metricmulti} gives
\begin{equation}
-\FF R
=
8\pi T^{(m)}
-\mathcal{B}_{AB}\nabla_a\phi^A\nabla^a\phi^B
-3\Box\FF
-2U.
\label{eq:trace_multi}
\end{equation}

Equations~\eqref{eq:metricmulti}-\eqref{eq:trace_multi} are the exact field equations needed below. Their structure already anticipates the thermodynamic interpretation. The nonminimal coupling enters through the Hessian sector generated by $\FF$, while the scalar fields also contribute through the field-space kinetic term controlled by $\mathcal{B}_{AB}$ and through the potential. Relative to the one-field case, the essential new feature is that the kinetic sector now carries a genuinely multi-channel structure in field space.

For the thermodynamic interpretation it is useful to rewrite Eq.~\eqref{eq:metricmulti} in Einstein-like form:
\begin{equation}
G_{ab}
=
8\pi
\left(
\frac{T^{(m)}_{ab}}{\FF}
+T^{(g)}_{ab}
\right),
\label{eq:einsteinlike_multi}
\end{equation}
where the geometric effective stress tensor is
\begin{align}
T^{(g)}_{ab}&=\frac{1}{8\pi\FF}\left[\nabla_a\nabla_b\FF-g_{ab}\Box\FF-\frac12 U\,g_{ab}\right.\nonumber\\
&\left.+\mathcal{B}_{AB}\left(\nabla_a\phi^A\nabla_b\phi^B-\frac12g_{ab}\nabla_c\phi^A\nabla^c\phi^B\right)
\right].
\label{eq:Tg_multi}
\end{align}
\section{Covariant $1+3$ decomposition in a general configuration: effective scalar fluid}
\label{sec:covariant}
With the theory specified, we now rewrite the geometric sector in the $1+3$ language needed for the effective-fluid and thermodynamic interpretations. Let $u^a$ be an arbitrary unit timelike vector field, with $u^a u_a=-1$. We then define the spatial projector
\begin{equation}
h_{ab}=g_{ab}+u_a u_b,
\end{equation}
whose covariant derivative decomposes as
\begin{equation}
\nabla_a u_b=-a_b u_a+\frac{1}{3}\Th h_{ab}+\sigma_{ab}+\omega_{ab},
\label{eq:ukinematics}
\end{equation}
where $\Th$ is the expansion scalar and $a^a$ is the 4-acceleration, defined as
\begin{equation}
\Th\equiv \nabla_a u^a,
\qquad
a^a\equiv u^b\nabla_b u^a,
\end{equation}
and $\sigma_{ab}$ and $\omega_{ab}$ are the shear and vorticity defined as
\begin{equation}
\sigma_{ab}
\equiv
h_{(a}{}^{c}h_{b)}{}^{d}\nabla_c u_d
-\frac13\Theta h_{ab},
\end{equation}
and
\begin{equation}
\omega_{ab}
\equiv
h_{[a}{}^{c}h_{b]}{}^{d}\nabla_c u_d.
\end{equation}
where round brackets denote symmetrization,
\begin{equation}
X_{(ab)}
\equiv
\frac12\left(X_{ab}+X_{ba}\right),
\end{equation}
while square brackets denote antisymmetrization,
\begin{equation}
X_{[ab]}
\equiv
\frac12\left(X_{ab}-X_{ba}\right).
\end{equation}

Equivalently, one can also write
\begin{equation}
\sigma_{ab} = \left( h_{(a}{}^{c}h_{b)}{}^{d} -\frac13 h_{ab}h^{cd} \right)\nabla_c u_d,
\end{equation}
\begin{equation}
\omega_{ab}
=
h_{[a}{}^{c}h_{b]}{}^{d}\nabla_c u_d,
\end{equation}
satisfying the properties
\begin{equation}
\sigma_{ab}=\sigma_{ba}, \qquad \sigma^a{}_a=0, \qquad u^a\sigma_{ab}=0,
\end{equation}
and
\begin{equation}
\omega_{ab}=-\omega_{ba},
\qquad
u^a\omega_{ab}=0.
\end{equation}

For any scalar $\varphi$ we define
\begin{equation}
\dot{\varphi}\equiv u^a\nabla_a \varphi,
\qquad
\D_a \varphi\equiv h_a{}^b\nabla_b \varphi,
\qquad
\D^2 \varphi\equiv \D_a\D^a \varphi\,,
\end{equation}
giving the following relations that will be used throughout the work:
\begin{align}
\nabla_a \varphi&=-\dot{\varphi}u_a+\D_a \varphi,
\label{eq:graddecomp}
\\
\Box \varphi&=-\ddot{\varphi}-\Th\dot{\varphi}+\D^2\varphi+a^a\D_a \varphi,
\label{eq:boxdecomp}
\\
u^a u^b\nabla_a\nabla_b \varphi&=\ddot{\varphi}-a^a\D_a \varphi,
\label{eq:uuderiv}
\\
h^{ab}\nabla_a\nabla_b \varphi&=-\Th\dot{\varphi}+\D^2\varphi.
\label{eq:spatderiv}
\\
h_a{}^b u^c \nabla_c \nabla_b \varphi
&=
h_a{}^b u^c \nabla_c(\D_b \varphi)-\dot \varphi\, a_a \\
&=\D_a\dot{\varphi} - \left(\frac{1}{3}\Th h_a{}^c+\sigma_a{}^c + \omega_a{}^c\right) \D_c \varphi.
\label{eq:keyidentity2}
\end{align}

We also define the iterated projected derivative by
\begin{equation}
\D_a\D_b \varphi\equiv h_a{}^c h_b{}^d \nabla_c\!\left(h_d{}^e\nabla_e \varphi\right).
\label{eq:DDdef}
\end{equation}
A direct computation then gives
\begin{equation}
\D_a\D_b \varphi
=
h_a{}^c h_b{}^d \nabla_c \nabla_d \varphi
+\dot \varphi\left(\frac13\Th h_{ab}+\sigma_{ab}+\omega_{ab}\right),
\label{eq:DDidentity}
\end{equation}
and therefore
\begin{equation}
h_{\langle a}{}^c h_{b\rangle}{}^d \nabla_c \nabla_d \varphi
=
\D_{\langle a}\D_{b\rangle}\varphi-\dot \varphi\,\sigma_{ab}.
\label{eq:PSTFidentity}
\end{equation}
where we denote by angle brackets the projected, symmetric, trace-free (PSTF) part of a rank-two tensor with respect to $u^a$. For any tensor $X_{ab}$,
\begin{equation}
X_{\langle ab\rangle}
\equiv
\left[
h_{(a}{}^{c}h_{b)}{}^{d}
-\frac13 h_{ab}h^{cd}
\right]X_{cd}.
\label{eq:PSTFdef}
\end{equation}

In particular, $X_{\langle ab\rangle}$ is orthogonal to $u^a$ on both indices, symmetric, and trace-free:
\begin{equation}
X_{\langle ab\rangle}=X_{\langle ba\rangle},
\qquad
u^a X_{\langle ab\rangle}=0,
\qquad
h^{ab}X_{\langle ab\rangle}=0.
\end{equation}

Relative to any chosen congruence $u^a$, in the $1+3$ decomposition, the geometric effective stress tensor~\eqref{eq:Tg_multi} can be written in the form of an imperfect fluid 
\begin{equation}
T^{(g)}_{ab}
=
\rho_g u_a u_b
+p_g h_{ab}
+2q^{(g)}_{(a}u_{b)}
+\pi^{(g)}_{ab},
\label{eq:taufluid_multi}
\end{equation}
where
\begin{align}
\rho_g&\equiv u^au^bT^{(g)}_{ab},
\\
p_g&\equiv \frac13 h^{ab}T^{(g)}_{ab},
\\
q^{(g)}_a&\equiv -h_a{}^c u^d T^{(g)}_{cd},
\\
\pi^{(g)}_{ab}&\equiv h_{\langle a}{}^c h_{b\rangle}{}^d T^{(g)}_{cd}.
\end{align}

Using \eqref{eq:graddecomp}-\eqref{eq:keyidentity2} in combination with the previous definitions yields then the effective fluid quantities:
\begin{align}
\rho_g
&=
\frac{1}{8\pi\FF} \left[ -\Theta\dot\FF+\D^2\FF +\frac12\mathcal{B}_{AB} \left( \dot\phi^A\dot\phi^B+\D_a\phi^A\D^a\phi^B \right) \right. \nonumber \\
&\left.+\frac{U}{2}
\right],
\label{eq:rhog_multi}
\\
p_g
&=
\frac{1}{8\pi\FF}
\Bigg[
\ddot\FF+\frac23\Theta\dot\FF-\frac23\D^2\FF-a^a\D_a\FF
\nonumber\\
& +\frac12\mathcal{B}_{AB}\dot\phi^A\dot\phi^B -\frac16\mathcal{B}_{AB}\D_a\phi^A\D^a\phi^B -\frac{U}{2}
\Bigg],
\label{eq:pg_multi}
\\
q^{(g)}_a
&=
\frac{1}{8\pi\FF}
\left[
\dot\FF\,a_a
-h_a{}^b u^c\nabla_c\D_b\FF
-\mathcal{B}_{AB}\dot\phi^A\D_a\phi^B
\right],
\label{eq:qg_multi}
\\
\pi^{(g)}_{ab}
&=
\frac{1}{8\pi\FF}
\left[
\D_{\langle a}\D_{b\rangle}\FF-\dot\FF\,\sigma_{ab}
+\mathcal{B}_{AB}\D_{\langle a}\phi^A\D_{b\rangle}\phi^B
\right].
\label{eq:pig_multi_multi}
\end{align}

These expressions already hint at the thermodynamic structure of the multi-field theory: the imperfect-fluid variables are naturally split into a coupling sector governed by $\FF$ and a residual field-space sector governed by the scalar gradients weighted by $\mathcal{B}_{AB}$.

\section{Eckart thermodynamics in a generic frame}
\label{sec:thermal}

A central pillar of the thermodynamic interpretation of scalar-tensor gravity is its connection with Eckart's first-order theory of relativistic dissipation~\cite{Eckart:1940te}. The key point is not merely that the scalar sector can be written as an effective imperfect fluid, but that its effective heat flux $q_a$, bulk pressure $\Pi$, and anisotropic stresses $\pi_{ab}$ given by
\begin{align}
\Pi &= -\zeta\,\Theta,\label{eq:eckartP} \\
q_a^{\text{Eckart}}&=-K\left(\D_a T+Ta_a\right)\label{eq:eckart}, \\
\pi_{ab}^{\text{Eckart}} &= -2\eta\,\sigma_{ab}\label{eq:eckartp},
\end{align}
where $\zeta$ is the bulk-viscosity coefficient, $K$ is the thermal conductivity, and $\eta$ is the shear-viscosity coefficient, can be compared directly with the constitutive variables of relativistic irreversible thermodynamics. These quantities are not matter thermodynamic variables. In particular, $T$ should not be identified with the temperature measured by a matter thermometer unless an additional physical prescription specifies thermal contact or equilibration between the matter sector and the effective geometric medium. The same qualification applies to the entropy current and entropy production introduced below. They are first-order effective quantities associated with the matched geometric medium. They are not microscopic entropy variables, and they are not derived from a causal second-order theory.

{Moreover, it is the constitutive identification that gives physical meaning to the effective-fluid description: once the gravitational scalar sector is matched to the Eckart form, one can introduce transport coefficients, an effective temperature, and a notion of thermal evolution relative to the Einstein limit \cite{Faraoni:2021lfc,Faraoni:2021jri,Faraoni:2022gry,Giardino:2022sdv,Faraoni:2023hwu,Houle:2024sxs,Miranda:2024dhw,Giardino:2023ygc,Faraoni:2025alq}}. In this sense, the Eckart correspondence is not a secondary reinterpretation of the field equations, but the step that turns the effective fluid into a genuine thermodynamic medium and underlies the interpretation of General Relativity as a zero-temperature equilibrium state. At the same time, this connection should be handled with some care: as is well known, first-order relativistic thermodynamics has limitations at the level of causality and stability, and in modified gravity the existence of a theory heat flux does not automatically guarantee the existence of a bona fide Eckart temperature in an arbitrary frame \cite{Maartens:1996vi,Hiscock:1985zz,Pereira:2025dmk}. It is therefore the constitutive matching itself that becomes the decisive step in any first-order thermodynamic analysis of scalar-tensor theories. The strategy of this section is the following. We first isolate the exact structure of the theory heat flux in a generic frame and identify the pair $(\chi,W_a)$ that would match an Eckart description if a local temperature exists. We then separate the mathematical integrability problem from its physical interpretation: the derivation establishes the exact form of $q_a^{(g)}$, while the subsequent discussion clarifies when $W_a$ can be interpreted as a genuine temperature-gradient sector. We conclude by commenting on the more restrictive matching conditions in the anisotropic-stress and bulk sectors.
\subsection{Heat-flux matching in a generic frame}
We begin with the heat flux $q_a$, since it underlies the thermodynamic interpretation of GR as a zero-temperature equilibrium configuration.~\cite{Faraoni:2025alq}. 

A point that is easy to miss, and that is often hidden in the one-field literature by a convenient frame choice, is the origin of the explicit spatial-gradient terms in the heat flux. In the one-field scalar-tensor case (excluding more general derivative theories such as Horndeski), one usually adopts the scalar-comoving frame $u_a\propto \nabla_a\phi$ whenever $\nabla_a\phi$ is timelike. Then $\D_a\phi=0$, the explicit spatial-gradient terms disappear, and the heat flux reduces to an Eckart-type inertial piece proportional to $a_a$. In the present multi-field theory this simplification is generically unavailable. One may choose a frame comoving with a given scalar field $\phi^I$, but such a choice does not in general make the remaining gradients vanish. Unless all scalar gradients are mutually aligned, as will be seen later, one cannot simultaneously impose $\D_a\phi^A=0$ for every field. This geometric reason is expected to lead to explicit spatial-gradient terms that survive in the total heat flux~\cite{Miranda:2022uyk} making the system behave as a genuine multi-component medium rather than as a single Eckart fluid.

The correct question in the present multi-field setting is therefore not whether $q_a^{(g)}$ contains an acceleration term--it does--but whether the \emph{entire} $q_a^{(g)}$ can be cast in the form \eqref{eq:eckart}. Moreover, this possible problem is already present in one-field cases as analyzed in~\cite{Pereira:2025dmk,Giusti:2021sku} where the obstruction arises because the coupling function to curvature depends on derivatives of the scalar-field.

Starting from Eq.~\eqref{eq:qg_multi}, using the identity~\eqref{eq:keyidentity2}, the heat flux can be rewritten as
\begin{align}
q_a^{(g)}
&=
\frac{1}{8\pi\FF}
\Bigg[
-\D_a\dot\FF
+\left(
\sigma_a{}^b+\omega_a{}^b+\frac13\Theta h_a{}^b
\right)\D_b\FF
\nonumber\\
&
-\mathcal{B}_{AB}\dot\phi^A\D_a\phi^B
\Bigg].
\label{eq:qg_multi_expanded}
\end{align}

Equation~\eqref{eq:qg_multi_expanded} shows that, in a generic frame, the theory heat flux does not isolate an inertial term uniquely by coefficient matching, because the acceleration contribution contained in
$h_a{}^b u^c \nabla_c \D_b\FF$ exactly cancels the explicit $\dot\FF\,a_a$ term. Therefore, one should not identify $KT$ in a generic frame by matching the coefficient of $a_a$ alone.

Instead, on the branch $\dot\FF\neq 0$, it is convenient to introduce the exact decomposition
\begin{equation}
q_a^{(g)}=-\chi\left(a_a+W_a\right),
\label{eq:qg_decomp_generic}
\end{equation}
with
\begin{equation}
\chi\equiv -\frac{\dot\FF}{8\pi\FF},
\label{eq:chi_generic_def}
\end{equation}
and
\begin{align}
W_a
&\equiv
-\frac{1}{\dot\FF}
\Bigg[
\D_a\dot\FF
-\left(
\sigma_a{}^b+\omega_a{}^b+\frac13\Theta h_a{}^b
\right)\D_b\FF
\nonumber\\
&
+\dot\FF\,a_a
+\mathcal{B}_{AB}\dot\phi^A\D_a\phi^B
\Bigg].
\label{eq:Wdef_multi_corrected}
\end{align}
Indeed, substituting Eqs.~\eqref{eq:chi_generic_def} and \eqref{eq:Wdef_multi_corrected} into Eq.~\eqref{eq:qg_decomp_generic} reproduces Eq.~\eqref{eq:qg_multi} exactly.

Equations~\eqref{eq:qg_decomp_generic}--\eqref{eq:Wdef_multi_corrected} are purely exact identities. Their thermodynamic content depends on whether $W_a$ can be realized as a projected logarithmic temperature gradient; this is the constitutive question addressed next.

With this structure in place, the central thermodynamic issue in determining whether the geometric heat flux $q_a^{(g)}$ admits an Eckart interpretation is whether the spatial covector $W_a$ can be expressed as the projected gradient of a scalar temperature field. That is, whether there exists a scalar field $T$ such that
\begin{equation}
\D_a\ln T=W_a.
\label{eq:Tsolve_multi}
\end{equation}
If such a scalar exists, then the theory heat flux admits the Eckart form
\begin{equation}
q_a^{(g)}=-KT\left(a_a+\D_a\ln T\right)
=-K\left(\D_aT+Ta_a\right),
\label{eq:eckart_match_generic}
\end{equation}
with
\begin{equation}
KT=\chi=-\frac{\dot\FF}{8\pi\FF}.
\label{eq:KTmatch_multi}
\end{equation}

Equation~\eqref{eq:Tsolve_multi} is a nontrivial integrability problem. To see this, note first that Eq.~\eqref{eq:Tsolve_multi} is an overdetermined first-order system for the single scalar unknown $T$. Therefore, not every spatial covector $W_a$ can be written as $\D_a \ln T$. A necessary condition is obtained by applying $\D_{[a}$ to both sides: 
\begin{equation} 
\D_{[a}\D_{b]} \ln T = \D_{[a} W_{b]}. 
\label{eq:applycurl} 
\end{equation} 
with square brackets denoting antisymmetrization, so that 
\begin{equation} \D_{[a}W_{b]} \equiv \frac12\left(\D_aW_b-\D_bW_a\right). 
\end{equation} 

If the projected derivatives $\D_a$ behaved like ordinary derivatives on a three-dimensional hypersurface, one would conclude that the left-hand side vanishes identically. However, in the $1+3$ covariant formalism this is true only when the observer congruence is hypersurface-orthogonal. In general, the projected derivatives do not commute on scalars, and one has the standard commutator identity 
\begin{equation} 
\D_{[a}\D_{b]} f = -\omega_{ab}\dot f \label{eq:scalarcurl_app} 
\end{equation} 
for any scalar $f$, where $\omega_{ab}$ is the vorticity tensor of the congruence and $\dot f \equiv u^c \nabla_c f$ \cite{Tsagas:2007yx,Ellis:1971pg}. Applying Eq.~\eqref{eq:scalarcurl_app} to $f=\ln T$ and comparing with Eq.~\eqref{eq:applycurl} gives 
\begin{equation} 
\D_{[a}W_{b]} = -\omega_{ab}\,\dot{\left(\ln T\right)}. \label{eq:compatibility_app} 
\end{equation} 

Equation~\eqref{eq:compatibility_app} is therefore not an additional assumption but the necessary compatibility condition for the local existence of a temperature field satisfying Eq.~\eqref{eq:Tsolve_multi}. When $\omega_{ab}\neq0$, the observers' local rest spaces do not fit together into spacelike hypersurfaces, and the projected derivative $\D_a$ does not behave as an ordinary spatial gradient operator. In that case, the ``curl'' of a projected gradient need not vanish, and the existence of a temperature field is constrained by the vorticity of the chosen congruence. Only in the irrotational case, 
\begin{equation} 
\omega_{ab}=0, \label{eq:irrot_app} 
\end{equation} 
does Eq.~\eqref{eq:compatibility_app} reduce to the familiar closed condition \begin{equation} 
\D_{[a}W_{b]}=0, \label{eq:compatibility_irrot_app} 
\end{equation} 
which is especially relevant below when the analysis is specialized to the natural scalar-comoving frames. Thus, the existence of a matching between the multi-scalar theory heat flux and an Eckart heat flux consists of knowing whether Eq.~\eqref{eq:compatibility_app}, or Eq.~\eqref{eq:compatibility_irrot_app} in an irrotational frame, can be satisfied for the configuration under consideration. Moreover, if in a given configuration these conditions are satisfied, then $W_a$ determines the spatial dependence of the local temperature field through Eq.~\eqref{eq:Tsolve_multi}. Once $T$ is reconstructed, the conductivity follows from the matched product 
\begin{equation} 
\chi \equiv KT \label{eq:Xidef_app} 
\end{equation} 
as 
\begin{equation} 
K = \frac{\chi}{T} = -\frac{\dot{\FF}}{8\pi \FF\,T}. \label{eq:Keff} 
\end{equation} 

However, this reconstruction is not unique: the solution for $T$ is fixed only up to a multiplicative factor with vanishing spatial gradient, and the same residual freedom is inherited by $K$ unless an additional constitutive prescription is imposed. This point is also reflected in the evolution along the flow. Since 
\begin{equation} 
\dot{\left(\ln T\right)} = \dot{\left(\ln \chi\right)} - \dot{\left(\ln K\right)}, \label{eq:lnTvslnXi} 
\end{equation} 
the compatibility condition~\eqref{eq:compatibility_app} is not closed unless one specifies how $K$ evolves. For example, if $K$ is assumed to be constant along the flow, then $\dot{\left(\ln T\right)}=\dot{\left(\ln \chi\right)}$ and the compatibility relation can be expressed entirely in terms of the theory variables. Without such an input, Eq.~\eqref{eq:compatibility_app} should be understood as the exact necessary condition for the existence of a local Eckart temperature, but not yet as a closed evolution equation for it.

\subsection{Anisotropic-stress and bulk-viscous sectors}
\label{sec:shearbulk}

The heat-flux matching discussed above is sufficient for the temperature-based thermodynamic analysis developed in this work. One may nevertheless ask whether the remaining dissipative sectors of the generalized multi-scalar medium can also be written in Eckart form. The answer is more restrictive than in the heat-flux sector, because here one must match not only scalars but also the full tensorial structure of the effective source. Starting with the shear, comparing Eq.~\eqref{eq:pig_multi_multi} with Eq.~\eqref{eq:eckartp} one can see that the match demands that the full projected trace-free tensor appearing in Eq.~\eqref{eq:pig_multi_multi} be proportional to the shear tensor of the chosen congruence. In other words, the multi-scalar medium admits an Eckart shear-viscous interpretation only on those branches for which the Hessian and scalar-gradient contributions align with $\sigma_{ab}$. A general condition for the matching to be possible is given by
\begin{equation}
\mathcal{S}_{ab}
\equiv
\D_{\langle a}\D_{b\rangle}\FF
+\mathcal{B}_{AB}\D_{\langle a}\phi^A\D_{b\rangle}\phi^B
=
\Lambda\,\sigma_{ab}
\label{eq:Sdef_multi}
\end{equation}
for some scalar $\Lambda$. In that case,
\begin{equation}
\eta_{\rm eff}
=
\frac{\dot\FF-\Lambda}{16\pi\FF}.
\label{eq:etageneric_multi}
\end{equation}
and if $\sigma_{ab}=0$, the Eckart match requires instead $\pi^{(g)}_{ab}=0$. This matching condition is not trivial and restricts the configuration space. 

The bulk-viscous sector is subtler. Unlike the heat flux and anisotropic stress, the identification of a bulk viscous pressure requires first a split
\begin{equation}
p_g=P_{\rm eq}+\Pi_g,
\label{eq:peqPi}
\end{equation}
and only after such a split has been chosen can one test whether $\Pi_g=-\zeta_{\rm eff}\,\Theta$. This means that the bulk-viscous interpretation is partly conventional: the imperfect-fluid decomposition fixes $p_g$, but it does not uniquely determine which part of it should be regarded as equilibrium pressure and which part as nonequilibrium bulk pressure. At this stage two natural choices are available.

One may adopt the same economical convention often used in the one-field literature and absorb the entire isotropic sector into an effective equilibrium pressure. Then
\begin{equation}
\Pi=0,
\qquad
\zeta=0.
\label{eq:zetazero}
\end{equation}

This choice has the advantage of isolating the genuinely dissipative content in the heat-flux and anisotropic-stress sectors alone.

Alternatively, one may isolate explicitly the term proportional to the expansion scalar. For a generic congruence $u^a$, using Eq.~\eqref{eq:pg_multi}, one may write
\begin{equation}
p_g
=
\frac{1}{8\pi\FF}
\left[
\mathcal{P}[u]+\frac23\Theta\dot\FF
\right],
\end{equation}
with
\begin{align}
\mathcal{P}[u]
&\equiv
\ddot\FF-\frac23\D^2\FF-a^a\D_a\FF
+\frac12\mathcal{B}_{AB}\dot\phi^A\dot\phi^B
\nonumber\\
&
-\frac16\mathcal{B}_{AB}\D_a\phi^A\D^a\phi^B
-\frac{U}{2}.
\end{align}

Hence, if one adopts the minimal split
\begin{equation}
P_{\rm eq}^{\rm(min)}
=
\frac{\mathcal{P}[u]}{8\pi\FF},
\end{equation}
then
\begin{equation}
\Pi_g^{\rm(min)}
=
\frac{\dot\FF}{12\pi\FF}\,\Theta,
\qquad
\zeta_{\rm eff}^{\rm(min)}
=
-\frac{\dot\FF}{12\pi\FF}.
\label{eq:zetamin_multi}
\end{equation}

\section{Natural frame specializations}
\label{sec:frames}

The generic-$u^a$ formulas above are the fundamental statements of the theory. Special frame choices are then optional tools for interpretation. In a genuinely multi-field theory there is, in general, no unique preferred scalar-comoving frame~\cite{Miranda:2022uyk}. 

In the present multi-field theory, two natural classes of comoving frames can be distinguished: frames comoving with an individual scalar field and the frame comoving with the effective curvature coupling $\FF$. More precisely, for any selected scalar $\phi^I$ whose gradient is timelike, one may define the $\phi^I$-comoving frame by
\begin{equation}
u_a^{(I)}
\equiv
\frac{\nabla_a\phi^I}{\sqrt{-\nabla_b\phi^I\nabla^b\phi^I}},
\qquad
\nabla_a\phi^I\nabla^a\phi^I<0,
\label{eq:uI_multi}
\end{equation}
while, whenever the gradient of the coupling function is timelike, one may define the coupling-comoving frame by
\begin{equation}
u_a^{(\FF)}
\equiv
\frac{\nabla_a\FF}{\sqrt{-\nabla_b\FF\nabla^b\FF}},
\qquad
\nabla_a\FF\nabla^a\FF<0.
\label{eq:uF_multi}
\end{equation}

In the language of~\cite{Miranda:2022uyk}, the first option is channel-comoving frames and the second is the average coupling frame. All of them are legitimate choices, although some may admit a more direct physical interpretation and provide a more convenient framework for the thermodynamic analysis as discussed below. Furthermore, a useful structural fact is that each scalar-comoving frame is automatically hypersurface-orthogonal and therefore irrotational wherever the defining gradient is timelike. Therefore the generic compatibility relation \eqref{eq:compatibility_app} collapses to a closed local curl condition in all the natural scalar-comoving frames. 

With this established, we will now determine the consequences for the thermodynamic construction once the generic formulas are specialized to each frame. The guiding principle is that the theory supplies $q_a^{(g)}[u]$, $\Pi[u]$, and $\pi^{(g)}_{ab}[u]$ for every congruence $u^a$, but the constitutive quantities $K[u]$, $T[u]$, $\zeta[u]$, and $\eta[u]$ exist only when the matching conditions of Section~\ref{sec:thermal} are satisfied in that frame.

\subsection{A selected-field comoving frame}

Take $u_a=u_a^{(I)}$ for some chosen scalar $\phi^I$. Then
\begin{equation}
\D_a^{(I)}\phi^I=0.
\end{equation}

In this frame the total heat flux becomes
\begin{equation}
q_a^{(g)}\Big|_{(I)}
=
\frac{1}{8\pi\FF}
\left[
\dot\FF a_a
-h_a{}^b u^c\nabla_c\D_b\FF
-\mathcal{B}_{AB}\dot\phi^A\D_a\phi^B
\right]_{(I)},
\label{eq:qgI_multi}
\end{equation}
where the superscript/subscript ${(I)}$ indicates that the corresponding quantity is defined with respect to the $\phi^I$-comoving frame. The same notation will be used throughout for the other frame.

Analyzing Eq.~\eqref{eq:qgI_multi}, one can see that unlike in the one-field case, a selected-field frame does not in general eliminate the full spatial structure of the heat flux. It merely removes the explicit spatial gradient of the chosen field, while the remaining fields still contribute through both $\D_a\FF$ and the field-space kinetic term.

The corresponding matching conditions for the heat flux are then
\begin{align}
\chi_{(I)}=K_{(I)}T_{(I)}
&=
-\frac{\dot\FF_{(I)}}{8\pi\FF},
\\
\D_{[a}W^{(I)}_{b]}
&=0,
\end{align}
with
\begin{align}
W_a^{(I)}
&=
-\frac{1}{\dot\FF_{(I)}}
\Bigg[
\D_a^{(I)}\dot\FF
-\left(
\sigma_a{}^b+\frac13\Theta h_a{}^b
\right)_{(I)}
\D_b^{(I)}\FF
\nonumber\\
&+\dot{\FF} a_a
+\mathcal{B}_{AB}\dot\phi^A\D_a\phi^B\Big|_{(I)}
\Bigg].
\end{align}

For the anisotropic-stress sector, the $\phi^I$-comoving frame yields
\begin{align}
\pi^{(g)}_{ab}\Big|_{(I)}&= \frac{1}{8\pi \FF}\left(\D_{\langle a}^{(I)}\D_{b\rangle}^{(I)}\FF-\dot \FF_{(I)}\,\sigma^{(I)}_{ab}\right) \nonumber \\
&+\frac{1}{8\pi \FF}\,\mathcal{B}_{\hat A\hat B}\,\D_{\langle a}^{(I)}\phi^{\hat A}\,\D_{b\rangle}^{(I)}\phi^{\hat B},
\label{eq:piIframeMulti}
\end{align}
where hatted field-space indices run over the scalars other than $\phi^I$. Hence the Eckart shear-viscous match
\begin{equation}
\pi^{(g)}_{ab}\Big|_{(I)}
=
-2\eta_{(I)}\,\sigma^{(I)}_{ab}
\label{eq:pimatchIframeMulti}
\end{equation}
exists if and only if
\begin{equation}
\D_{\langle a}^{(I)}\D_{b\rangle}^{(I)}\FF
+
\mathcal{B}_{\hat A\hat B}\,
\D_{\langle a}^{(I)}\phi^{\hat A}\,
\D_{b\rangle}^{(I)}\phi^{\hat B}
=
\Lambda_{(I)}\,\sigma^{(I)}_{ab},
\label{eq:piCondIframeMulti}
\end{equation}
for some scalar $\Lambda_{(I)}$, in which case
\begin{equation}
\eta_{(I)}
=
\frac{\dot \FF_{(I)}-\Lambda_{(I)}}{16\pi \FF}.
\label{eq:etaIframeMulti}
\end{equation}

For the isotropic sector one has
\begin{align}
p_g\Big|_{(I)}
&=
\frac{1}{8\pi \FF}
\Bigg[
\ddot \FF_{(I)}
+\frac{2}{3}\Theta_{(I)}\dot \FF_{(I)}
-\frac{2}{3}\D^{2}_{(I)}\FF
-a^a_{(I)}\D_a^{(I)}\FF
\nonumber\\
&
+\frac{1}{2}\mathcal{B}_{AB}\,
\dot\phi^A_{(I)}\dot\phi^B_{(I)}
-\frac{1}{6}\mathcal{B}_{\hat A\hat B}\,
\D_a^{(I)}\phi^{\hat A}\D^a_{(I)}\phi^{\hat B}
-\frac{U}{2}
\Bigg].
\label{eq:pIframeMulti}
\end{align}
A minimal bulk split then isolates the explicit expansion term:
\begin{equation}
P^{\rm(min)}_{(I)}
\equiv
p_g\Big|_{(I)}
-
\frac{\dot \FF_{(I)}}{12\pi \FF}\,\Theta_{(I)},
\label{eq:peqIframeMulti}
\end{equation}
so that
\begin{equation}
\Pi^{\rm(min)}_{(I)}
=
\frac{\dot \FF_{(I)}}{12\pi \FF}\,\Theta_{(I)},
\quad
\zeta^{\rm(min)}_{(I)}
=
-\frac{\dot \FF_{(I)}}{12\pi \FF}.
\label{eq:zetaIframeMulti}
\end{equation}
As in the other frames, one may instead adopt the economical convention
\begin{equation}
\Pi_{(I)}=0.
\end{equation}

\subsection{Coupling/averaging-comoving frame}

Take $u_a=u_a^{(\FF)}$ whenever $\nabla_a\FF$ is timelike. Then $\D_a\FF=0$, and Eq.~\eqref{eq:qg_multi} reduces to
\begin{equation}
q_a^{(g)}\Big|_{(\FF)}
=
\frac{\dot\FF}{8\pi\FF}\,a_a^{(\FF)}
-\frac{1}{8\pi\FF}\,
\mathcal{B}_{AB}\dot\phi^A\D_a\phi^B\Big|_{(\FF)}.
\label{eq:qgF_multi}
\end{equation}
Hence the matching conditions are
\begin{align}
\chi_{\FF}=K_{\FF}T_{\FF}
&=
-\frac{\dot\FF}{8\pi\FF},
\label{eq:XiF_multi}
\\
\D_{[a}W^{(\FF)}_{b]}
&=0,
\label{eq:Fframematch_multi}
\end{align}
with
\begin{equation}
W_a^{(\FF)}=-\frac{1}{\dot\FF}\,
\mathcal{B}_{AB}\dot\phi^A\D_a\phi^B.
\label{eq:WF_multi}
\end{equation}

This expression makes the physical meaning of $W_a^{(\FF)}$ explicit. It is the residual thermal-gradient sector that survives after moving to the frame comoving with the effective coupling, and it vanishes only when the remaining scalar directions fail to generate a net field-space flux across the local rest spaces. Equivalently, $W_a^{(\FF)}$ measures the multi-field misalignment between the coupling direction selected by $\FF$ and the kinetic/spatial scalar structure selected by $\mathcal{B}_{AB}\dot\phi^A\D_a\phi^B$. When $W_a^{(\FF)}=0$ the heat flux is purely inertial, as in the usual one-field picture; when $W_a^{(\FF)}\neq0$, the thermal description is genuinely multi-field and cannot be reduced to $\chi_{\FF}$ alone.

Furthermore, even in the frame comoving with the effective gravitational coupling $\FF$, the remaining scalar sector generally does not disappear. The full heat flux takes an Eckart form only when the residual multi-field contribution can be absorbed into the gradient of a single effective temperature, so that the existence of an Eckart temperature in this frame is again controlled by the corresponding local integrability condition~\eqref{eq:Fframematch_multi}. Hence, the departure from a purely inertial heat flux is governed by the contraction of the field-space kinetic tensor with the time derivative and spatial gradient of the scalar multiplet. Moreover, in contrast with the other frame, the integrability condition is simpler. Indeed, there is one immediate class of solutions of Eq.~\eqref{eq:Fframematch_multi},
\begin{equation}\label{eq:inertial2class_multi}
\D_a\phi^A=0 \quad \forall A,
\end{equation}
leading to
\begin{equation}
W_a^{(\FF)}=0
\qquad \Longrightarrow \qquad
\D_a T_{\FF}=0.
\end{equation}

This solution states that all the remaining scalar fields are spatially homogeneous with respect to the congruence $u^a_{(\FF)}$. Since $\D_a^{(\FF)}\FF=0$ by construction, this implies that the gradients of $\FF$ and of all the active scalar fields are parallel to the same timelike congruence, so the system is driven into a fully aligned scalar configuration. This is realized, in particular, in homogeneous cosmology and will be studied ahead. Furthermore, it is important to note that the condition
\begin{equation}
\dot{\phi}^A_{(\FF)}=0
\qquad \forall A
\label{eq:allphidotzero}
\end{equation}
although sufficient to imply
\begin{equation}
W_a^{(\FF)}=0,
\end{equation}
is not admissible inside the nondegenerate $\FF$-comoving frame. Indeed, since $\FF=\FF(\phi^A)$, one has
\begin{equation}
\dot{\FF}_{(\FF)}
=
\frac{\partial \FF}{\partial \phi^A}\,\dot{\phi}^A_{(\FF)},
\label{eq:Fdotchainrule}
\end{equation}
and therefore Eq.~\eqref{eq:allphidotzero} implies
\begin{equation}
\dot{\FF}_{(\FF)}=0.
\end{equation}
which collapses the frame. Thus Eq.~\eqref{eq:allphidotzero} should be interpreted not as a branch of the present transport system, but as a degenerate endpoint where the coupling-frame description itself ceases to be the appropriate one.

What is admissible, however, is the weaker possibility that the contraction $\mathcal{B}_{AB}\dot{\phi}^A_{(\FF)}\D_a^{(\FF)}\phi^B$ effectively involves only a subset of scalar directions because some entries of the field-space metric $\mathcal{B}_{AB}$ vanish. In particular, if the nonzero contributions reduce to
\begin{equation}
\mathcal{B}_{AB}\dot{\phi}^A_{(\FF)}\D_a^{(\FF)}\phi^B
=
\mathcal{B}_{CD}\dot{\phi}^C_{(\FF)}\D_a^{(\FF)}\phi^D,
\label{eq:reducedfieldspaceflux}
\end{equation}
with the indices $C,D$ running over a proper subset of the full scalar multiplet, then it is sufficient to impose conditions only on that subset in order to obtain
\begin{equation}
W_a^{(\FF)}=0.
\end{equation}
For example, one may have
\begin{equation}
\dot{\phi}^C_{(\FF)}=0
\qquad
\text{for the subset entering Eq.~\eqref{eq:reducedfieldspaceflux}},
\end{equation}
while other scalar fields remain dynamically active and continue to support
\begin{equation}
\dot{\FF}_{(\FF)}\neq0.
\end{equation}

Therefore the purely inertial condition does not require all scalar velocities to vanish; it requires only the vanishing of the specific field-space flux selected by the nonzero components of $\mathcal{B}_{AB}$.

Whenever either of the conditions holds, the temperature-gradient sector disappears and the total heat flux reduces to a purely inertial form,
\begin{equation}
q_a^{(g)}\propto a_a,
\end{equation}
so that the Eckart matching collapses to the homogeneous-temperature case, $\D_aT_{\FF}=0$. In this restricted sector one recovers the same inertial structure familiar from the one-field analysis, and the thermal interpretation again reduces to a $KT$-type description.

Beyond these simple branches, more general solutions with $W_a^{(\FF)}\neq0$ depend on the structure of the scalar multiplet and of the field-space metric $\mathcal{B}_{AB}$. A useful nontrivial class is obtained by requiring that
\begin{equation}\label{eq:nonClass_multi}
\mathcal{B}_{AB}\,\dot{\phi}^A
=
\dot{\FF}\,\partial_B\Psi(\FF,\phi^C),
\end{equation}
for some smooth scalar function $\Psi$. Substituting Eq.~\eqref{eq:nonClass_multi} into Eq.~\eqref{eq:WF_multi} gives
\begin{equation}
W_a^{(\FF)}
=
-\partial_B\Psi\,\D_a\phi^B.
\end{equation}
Since $\D_a^{(\FF)}\FF=0$, one also has
\begin{equation}
\D_a^{(\FF)}\Psi
=
\partial_B\Psi\,\D_a^{(\FF)}\phi^B,
\end{equation}
and therefore
\begin{equation}
W_a^{(\FF)}=-\D_a^{(\FF)}\Psi,
\qquad\Longrightarrow\qquad
\D_{[a}W_{b]}^{(\FF)}=0.
\end{equation}
Thus the residual multi-field contribution is compatible with an Eckart temperature even though it does not vanish.

Once the compatibility condition is satisfied, the theory heat flux $q_a^{(g)}$ can be matched to the general Eckart form. This yields a thermodynamic description that is more general than in the one-field case. In the usual scalar-comoving treatment of a single scalar field, the heat flux reduces to a purely inertial form, so that the product $KT$ is the only relevant thermal variable and measures the strength of that inertial channel. In the present multi-field setting, by contrast, the matched heat flux also contains the quantity $W_a$, which encodes the spatial temperature-gradient sector. Accordingly, whenever $W_a\neq0$, the thermodynamic analysis can no longer be reduced to a single scalar variable of $KT$ type; the natural framework is instead a coupled analysis for the pair $(KT,W_a)$, as will be developed in the following section.

The anisotropic stress in the coupling frame becomes
\begin{equation}
\pi^{(g)}_{ab}\Big|_{(\FF)}
=
-\frac{\dot\FF_{(\FF)}}{8\pi\FF}\,\sigma_{ab}^{(\FF)}
+\frac{1}{8\pi\FF}\,
\mathcal{B}_{AB}\D_{\langle a}^{(\FF)}\phi^A\D_{b\rangle}^{(\FF)}\phi^B,
\label{eq:pig_multiF_multi}
\end{equation}
so the Eckart match exists iff
\begin{equation}
\mathcal{B}_{AB}\D_{\langle a}^{(\FF)}\phi^A\D_{b\rangle}^{(\FF)}\phi^B
=
\Lambda_{(\FF)}\,\sigma_{ab}^{(\FF)},
\label{eq:piCondF_multi}
\end{equation}
in which case
\begin{equation}
\eta_{(\FF)}
=
\frac{\dot\FF_{(\FF)}-\Lambda_{(\FF)}}{16\pi\FF}.
\label{eq:etaF_multi}
\end{equation}

The isotropic sector becomes
\begin{align}
p_g\Big|_{(\FF)}
&=
\frac{1}{8\pi\FF}
\Bigg[
\ddot\FF_{(\FF)}+\frac23\Theta_{(\FF)}\dot\FF_{(\FF)}
+\frac12\mathcal{B}_{AB}\dot\phi^A\dot\phi^B
\nonumber\\
&
-\frac16\mathcal{B}_{AB}\D_a\phi^A\D^a\phi^B
-\frac{U}{2}
\Bigg]_{(\FF)},
\end{align}
so that the minimal bulk split is
\begin{equation}
P^{\rm(min)}_{(\FF)}
\equiv
p_g\Big|_{(\FF)}
-\frac{\dot\FF_{(\FF)}}{12\pi\FF}\,\Theta_{(\FF)},
\end{equation}
and
\begin{equation}\label{eq:PiFF}
\Pi^{\rm(min)}_{(\FF)}
=
\frac{\dot\FF_{(\FF)}}{12\pi\FF}\,\Theta_{(\FF)}
=
-\frac{2}{3}\chi_{\FF}\,\Theta_{(\FF)},
\end{equation}
\begin{equation}\label{eq:zetaFF}
\zeta^{\rm(min)}_{(\FF)}
=
-\frac{\dot\FF_{(\FF)}}{12\pi\FF}
=
\frac{2}{3}\chi_{\FF}.
\end{equation}

Among the natural scalar-comoving frames, the coupling frame is the most useful at the general multi-field level because it is the only one that universally isolates the coupling contribution across all dissipative sectors. In selected-field frames, the simplifications remain model dependent and one typically removes only one scalar gradient at a time. By contrast, the coupling frame organizes the imperfect-fluid variables around the single quantity that governs the Einstein-like coupling of the theory, while collecting the remaining multi-field effects into residual field-space terms. Physically, this is the frame whose observers are comoving with the effective gravitational coupling in the sense that $\D_a^{(\FF)}\FF=0$: they see the coupling vary only along their flow, not across their local rest spaces. Thermodynamically, this makes the coupling channel locally ``at rest'' and separates its purely temporal nonequilibrium from the genuinely multi-scalar residual sector that survives through the other scalar directions. In this sense, the $\FF$-frame isolates the inertial thermal behavior associated with the effective coupling while making explicit any additional temperature-gradient structure that cannot be removed by adapting the congruence to $\FF$. For these reasons, the coupling frame offers not only the most tractable setup, but also the most transparent first-order thermodynamic interpretation of the generic multi-field system.

\subsection{Aligned-gradient sector}

A particularly simple and physically instructive sector is obtained when all scalar gradients are everywhere parallel. More precisely, suppose that there exists a scalar field $\varphi$ and a set of scalar functions $\lambda^A$ such that
\begin{equation}
\nabla_a\phi^A=\lambda^A\,\nabla_a\varphi .
\label{eq:alignedgradients_multi}
\end{equation}
Equation~\eqref{eq:alignedgradients_multi} implies that the scalar fields are locally functionally dependent: whenever $\nabla_a\varphi\neq0$, each $\phi^A$ may be written locally as a function of $\varphi$, so that
\begin{equation}
\lambda^A=\frac{d\phi^A}{d\varphi}.
\label{eq:lambdadef_multi}
\end{equation}
Decomposing Eq.~\eqref{eq:alignedgradients_multi} using Eq.~\eqref{eq:graddecomp} immediately gives
\begin{equation}
\dot{\phi}^A=\lambda^A\dot{\varphi},
\qquad
\D_a\phi^A=\lambda^A\D_a\varphi.
\label{eq:alignedsplit_multi}
\end{equation}

Since $\FF=\FF(\phi^A)$, one also has
\begin{equation}
\nabla_a\FF=\FF_{,A}\nabla_a\phi^A
=
\Lambda\,\nabla_a\varphi,
\qquad
\Lambda\equiv \FF_{,A}\lambda^A,
\label{eq:Faligned_multi}
\end{equation}
and therefore, whenever the relevant gradients are timelike, all scalar-comoving observer fields and the $\FF$-comoving observer field coincide up to the sign fixed by future-directed normalization:
\begin{equation}
u_a^{(A)}=\mathrm{sgn}(\lambda^A)\,u_a^{(\varphi)},
\qquad
u_a^{(\FF)}=\mathrm{sgn}(\Lambda)\,u_a^{(\varphi)}.
\label{eq:alignedframes_multi}
\end{equation}
Thus, away from the degenerate branch $\Lambda=0$ (for which $\nabla_a\FF=0$ and the $\FF$-comoving frame ceases to exist), all natural scalar-comoving frames collapse to a single common scalar-comoving frame.

Choosing that common frame, one has
\begin{equation}
\D_a\phi^A=0,
\qquad
\D_a\FF=0,
\label{eq:alignedcomoving_multi}
\end{equation}
and consequently all explicitly spatial scalar-gradient contributions to the heat flux vanish. The geometric heat flux therefore reduces to
\begin{equation}
q_a^{(g)}=\frac{\dot{\FF}}{8\pi\FF}\,a_a.
\label{eq:qgaligned_multi}
\end{equation}

Thus, in the aligned-gradient sector and in the common scalar-comoving frame, the multi-scalar medium acquires the same purely inertial Eckart form as an effectively one-field scalar-tensor system,
\begin{equation}
q_a^{(g)}=-K_{\FF}T_{\FF}\,a_a,
\qquad
K_{\FF}T_{\FF}=-\frac{\dot{\FF}}{8\pi\FF}.
\label{eq:chialigned_multi}
\end{equation}

The same simplification occurs in the anisotropic-stress sector. Since $\D_a\phi^A=0$ and $\D_a\FF=0$, Eq.~\eqref{eq:pig_multi_multi} reduces to
\begin{equation}
\pi^{(g)}_{ab}
=
-\frac{\dot{\FF}}{8\pi\FF}\,\sigma_{ab}.
\label{eq:pigaligned_multi}
\end{equation}
Hence, in the aligned-gradient branch, the geometric anisotropic stress is automatically of Eckart form,
\begin{equation}
\pi^{(g)}_{ab}=-2\eta_{\FF}\,\sigma_{ab},
\qquad
\eta_{\FF}=\frac{\dot{\FF}}{16\pi\FF}.
\label{eq:etaaligned_multi}
\end{equation}
Accordingly, the aligned-gradient sector removes the generic tensorial obstruction to an Eckart interpretation of the anisotropic-stress sector.

Turning now to the isotropic scalar extracted from the $1+3$ decomposition, one finds
\begin{equation}
p_g
=
\frac{1}{8\pi\FF}
\left[
\ddot{\FF}
+\frac{2}{3}\Theta\dot{\FF}
+\frac{1}{2}\mathcal{B}_{AB}\dot{\phi}^A\dot{\phi}^B
-\frac{U}{2}
\right].
\label{eq:Pgaligned_multi}
\end{equation}
Isolating the term explicitly proportional to the expansion scalar and adopting the minimal split yields
\begin{equation}
P_g^{\rm(min)}
\equiv
\frac{1}{8\pi\FF}
\left[
\ddot{\FF}
+\frac{1}{2}\mathcal{B}_{AB}\dot{\phi}^A\dot{\phi}^B
-\frac{U}{2}
\right],
\label{eq:peqaligned_multi}
\end{equation}
and
\begin{equation}
\Pi_g^{\rm(min)}
=
\frac{\dot{\FF}}{12\pi\FF}\,\Theta.
\label{eq:Piviscaligned_multi}
\end{equation}
Using the Eckart relation $\Pi=-\zeta\,\Theta$, one obtains
\begin{equation}
\zeta_{\FF}^{\rm(min)}
=
-\frac{\dot{\FF}}{12\pi\FF}.
\label{eq:zetaaligned_multi}
\end{equation}

Therefore, in the aligned-gradient sector, the heat-flux and anisotropic-stress channels automatically admit an Eckart interpretation, while the bulk-viscous sector remains subject to the same constitutive ambiguity as in the general case. The special simplification of this branch is that all nontrivial dissipative coefficients collapse to algebraic combinations of the single common scalar-comoving variable $\dot{\FF}/(8\pi\FF)$.

\section{Transport system and multi-scalar relaxation toward GR}
\label{sec:transport}

We now specialize the thermodynamic analysis to the $\FF$-comoving frame and derive the corresponding transport system. In this section we will identify the variables that govern the coupling-frame thermodynamic description of the geometric medium and afterwards clarify which of these variables diagnose only the relaxation of the coupling channel and which are required to characterize the full multi-scalar approach to the GR sector.

In the one-field case, the unique inertial variable of $KT$ type is sufficient to describe both the heat sector and the approach to the GR equilibrium state. In the multi-scalar theory this coincidence is lost. Even after passing to the frame adapted to the effective coupling $\FF$, scalar directions orthogonal to $\FF_{,A}$ in field space may remain dynamically active and continue to contribute to the thermal state. As a result, the coupling-frame thermodynamics is no longer exhausted by a single inertial variable.

\subsection{Coupling-frame variables and their physical roles}
\label{subsec:variables}

Once the geometric heat flux has been matched to the Eckart law in the $\FF$-comoving frame, the heat flux has the form
\begin{equation}\label{eq:heatFF}
    q_a^{(g)}\Big|_{(\FF)} = -\chi_{\FF}\left(a_a^{(\FF)}+W_a^{(\FF)}\right),
\end{equation}

Hence, the genuine thermodynamic variables of the effective medium are
\begin{equation}
\chi_{\FF}\equiv K_{\FF}T_{\FF},
\qquad
W_a^{(\FF)}\equiv \D_a^{(\FF)}\ln T_{\FF},
\label{eq:chiWdef}
\end{equation}
where $\chi_{\FF}$ governs the inertial coupling channel and $W_a^{(\FF)}$ measures the residual temperature-gradient sector across the local rest spaces. Thus $\chi_{\FF}$ controls the thermal response of the coupling channel along the flow, whereas $W_a^{(\FF)}$ describes the part of the thermal structure that survives after the frame has been adapted to $\FF$.

At first sight, $\chi_{\FF}$ looks like a single effective thermal variable. In a genuinely multi-scalar theory, however, it already contains nontrivial field-space structure. Since $\FF=\FF(\phi^A)$, one has
\begin{equation}
\dot{\FF}_{(\FF)}=\FF_{,A}\dot{\phi}^{A}_{(\FF)},
\end{equation}
and therefore
\begin{equation}
\chi_{\FF}
=
-\frac{\dot{\FF}_{(\FF)}}{8\pi\FF}
=
-\frac{\FF_{,A}\dot{\phi}^{A}_{(\FF)}}{8\pi\FF}.
\label{eq:chiFraw}
\end{equation}

This shows that the coupling thermal variable is not fundamental by itself: it is the coupling-direction projection of the scalar time-evolution amplitudes. This motivates introducing the field-space quantities
\begin{equation}
\chi^{A}\equiv -\frac{\dot{\phi}^{A}_{(\FF)}}{8\pi\FF},
\label{eq:chiAdef_transport}
\end{equation}
so that
\begin{equation}
\chi_{\FF}=\FF_{,A}\chi^A.
\label{eq:chiFfromchiA_transport}
\end{equation}

Hence $\chi_{\FF}$ is only the projection of the full scalar thermal vector $\chi^A$ along the coupling direction in field space. Furthermore, to identify the combinations that enter directly in the constitutive sector, it is also natural to define the lower-index quantity
\begin{equation}
\chi_A\equiv \mathcal{B}_{AB}\chi^B
=
-\frac{\mathcal{B}_{AB}\dot{\phi}^{B}_{(\FF)}}{8\pi\FF}.
\label{eq:chiAlowerdef_transport}
\end{equation}
that allows one to write Eq.~\eqref{eq:WF_multi} as
\begin{equation}
W_a^{(\FF)} = -\frac{\chi_A\,\D_a^{(\FF)}\phi^A}{\chi_{\FF}}.
\label{eq:WfromchiA}
\end{equation}

The distinction between $\chi^A$ and $\chi_A$ is physically meaningful. The variable $\chi^A$ encodes the full scalar velocity vector in thermal units and is therefore the natural quantity for discussing complete scalar relaxation. By contrast, $\chi_A$ is the kinetic-weighted thermal covector selected by the scalar kinetic matrix. It is this lower-index object that enters directly in the constitutive sector and, in particular, in the heat flux itself.

When $\mathcal{B}_{AB}$ is nondegenerate, it defines a genuine metric on field space and the two descriptions are equivalent:
\begin{equation}
\chi_A=\mathcal{B}_{AB}\chi^B,
\qquad
\chi^A=\mathcal B^{AB}\chi_B.
\label{eq:raiseLowerChi}
\end{equation}

When $\mathcal{B}_{AB}$ is degenerate, however, only the first relation remains meaningful. In that case $\chi_A$ is still well defined, but it no longer determines $\chi^A$ uniquely, because the kernel directions of $\mathcal{B}_{AB}$ are projected out. Accordingly, $\chi_A$ captures the thermal amplitudes seen by the kinetic sector, whereas $\chi^A$ retains the full information about the scalar time evolution.

If $\mathcal{B}_{AB}$ is nondegenerate, Eq.~\eqref{eq:chiFfromchiA_transport} may also be written in lower-index form as
\begin{equation}
\chi_{\FF}=\FF^{,A}\chi_A,
\qquad
\FF^{,A}\equiv \mathcal B^{AB}\FF_{,B},
\label{eq:chiFfromchiAlower_transport}
\end{equation}
that makes the hierarchy clear: the effective inertial variable $\chi_{\FF}$ is obtained from the full scalar thermal vector, or equivalently from its constitutive covector, by projection along the coupling direction.

Furthermore, the heat flux~\eqref{eq:heatFF} makes the roles of $\chi^A$ and $\chi_A$ especially transparent. Using the Eq.~\eqref{eq:chiFfromchiA_transport} and Eq.~\eqref{eq:WfromchiA} the heat flux can be recast in the form
\begin{align}
q_a^{(g)}\Big|_{(\FF)}= -\FF_{,A}\chi^A\,a_a^{(\FF)} +\chi_A\,\D_a^{(\FF)}\phi^A
\label{eq:qFfull}
\end{align}
where in the case of $\mathcal{B}_{AB}$ being nondegenerate
\begin{align}
q_a^{(g)}\Big|_{(\FF)} &= -\FF_{,A}\chi^A\, a_a^{(\FF)} +\mathcal{B}_{AB}\chi^B\,\D_a^{(\FF)}\phi^A \nonumber\\
&=-\FF^{,A}\chi_A\, a_a^{(\FF)} +\chi_A\,\D_a^{(\FF)}\phi^A.
\label{eq:qF_equivalentForms}
\end{align}

Equation~\eqref{eq:qFfull} shows that the two field-space quantities enter the heat sector in complementary ways. The upper-index thermal vector $\chi^A$ governs the inertial coupling part through its projection along $\FF_{,A}$, thereby producing the effective variable $\chi_{\FF}$. The lower-index quantity $\chi_A$, by contrast, governs the constitutive spatial part through its contraction with the projected scalar gradients $\D_a^{(\FF)}\phi^A$. Hence the effective coupling-frame heat flux is built simultaneously from the full scalar thermal vector and from its kinetic-weighted constitutive counterpart. The multi-scalar extension of the one-field picture is therefore more complex.

\subsection{Transport equations}
\label{subsec:transport_eqs}

We will now derive the evolution equations for $\chi_{\FF}$, $\chi^A$, $\chi_A$, and $W_a^{(\FF)}$. The first controls the inertial coupling channel, the second resolves the full scalar thermal vector, the third captures its kinetic-weighted constitutive projection, and the fourth measures the residual temperature-gradient sector that remains in the heat flux.

\subsubsection{$\chi_{\FF}$ sector}
As in the one-field case we will compute $\dot{\chi}_{\FF}$. We start by considering
\begin{align}
\dot\chi_{\FF}
&\equiv
u^a_{(\FF)}\nabla_a\left(-\frac{\dot\FF_{(\FF)}}{8\pi\FF}\right)
\nonumber\\
&=
-\frac{\ddot\FF_{(\FF)}}{8\pi\FF}
+\frac{\dot\FF_{(\FF)}^{\,2}}{8\pi\FF^2}
=
-\frac{\ddot\FF_{(\FF)}}{8\pi\FF}
+8\pi\,\chi_{\FF}^{\,2}.
\label{eq:chiFdot1}
\end{align}

Using identity \eqref{eq:boxdecomp} and $\D_a^{(\FF)}\FF=0$ allows to write $\Box \FF$ as
\begin{equation}
\Box\FF
=
-\ddot\FF_{(\FF)}-\Theta_{(\FF)}\dot\FF_{(\FF)}.
\label{eq:boxFcomoving}
\end{equation}
that allows to write Eq.~\eqref{eq:chiFdot1} as
\begin{align}
\dot\chi_{\FF}
&=
-\frac{1}{8\pi\FF}
\left(
-\Box\FF-\Theta_{(\FF)}\dot\FF_{(\FF)}
\right)
+8\pi\,\chi_{\FF}^{\,2}
\nonumber\\
&=
\frac{\Box\FF}{8\pi\FF}
-\Theta_{(\FF)}\chi_{\FF}
+8\pi\,\chi_{\FF}^{\,2}.
\label{eq:chiFtransport1}
\end{align}

Defining for convenience
\begin{equation}
\widetilde\Sigma_{\FF}
\equiv
\frac{\Box\FF}{8\pi\FF},
\label{eq:SigmaF}
\end{equation}
the coupling-channel transport equation~\eqref{eq:chiFtransport1} becomes
\begin{equation}
\dot\chi_{\FF}
=
8\pi\chi_{\FF}^{\,2}
-\Theta_{(\FF)}\chi_{\FF}
+\widetilde\Sigma_{\FF}.
\label{eq:chiFtransportSigma}
\end{equation}

Equation~\eqref{eq:chiFtransportSigma} is the natural multi-scalar generalization of the one-field coupling-temperature transport law. Its interpretation is unchanged: $\chi_{\FF}$ still governs the inertial thermal channel associated with the effective coupling $\FF$. What changes is the structure of the source term, which is no longer controlled by a single scalar degree of freedom but by the full multi-scalar dynamics.

When $\mathcal{B}_{AB}$ is nondegenerate, the source can be written more explicitly by contracting the scalar equations with $\FF^{,A}$ and using the trace of the metric field equations. Starting from
\begin{equation}
\Box\FF
=
\FF_{,A}\Box\phi^A
+
\FF_{,AB}\nabla_c\phi^A\nabla^c\phi^B,
\label{eq:boxFstart}
\end{equation}
and using the scalar equations in field-space covariant form, one obtains
\begin{equation}
\Box\FF
=
\mathcal{H}^{(\FF)}_{AB}\nabla_c\phi^A\nabla^c\phi^B
-\frac12\,\Xi_{\FF}R
+\frac12\,\FF^{,A}U_{,A},
\label{eq:boxFpretrace}
\end{equation}
where
\begin{equation}
\mathcal{H}^{(\FF)}_{AB}
\equiv
\nabla_A\nabla_B\FF
=
\FF_{,AB}-\Gamma^{C}{}_{AB}(\mathcal B)\FF_{,C}
\end{equation}
is the field-space covariant Hessian of the coupling function, while
\begin{equation}
\Xi_{\FF}
\equiv
\FF_{,A}\FF^{,A}
=
\mathcal B^{AB}\FF_{,A}\FF_{,B}
\end{equation}
is the field-space norm of its gradient. Moreover, using the trace of the metric field equation~\eqref{eq:trace_multi} one has
\begin{equation}
-\FF R
=
8\pi T^{(m)}
-\mathcal{B}_{AB}\nabla_c\phi^A\nabla^c\phi^B
-3\Box\FF
-2U,
\label{eq:trace_multi_transport}
\end{equation}
that gives
\begin{widetext}
\begin{equation}
\Box\FF=
\frac{
\mathcal{H}^{(\FF)}_{AB}\nabla_c\phi^A\nabla^c\phi^B
+\dfrac12\,\FF^{,A}U_{,A}
+\dfrac{4\pi\Xi_{\FF}}{\FF}T^{(m)}
-\dfrac{\Xi_{\FF}}{2\FF}\mathcal{B}_{AB}\nabla_c\phi^A\nabla^c\phi^B
-\dfrac{\Xi_{\FF}}{\FF}U
}{
1+\dfrac{3\Xi_{\FF}}{2\FF}
},
\label{eq:boxFexact}
\end{equation}
\end{widetext}
allowing to eliminate $\Box \FF$ in~\eqref{eq:chiFtransportSigma}.

At the formal level Eq.~\eqref{eq:chiFtransportSigma} is a Riccati-type transport equation along the $\FF$-comoving flow. If the geometric and matter quantities entering its coefficients are regarded as prescribed along each worldline, it takes the standard form $\dot y=a y^2+b y+c$, with
\begin{equation}
a=8\pi,
\qquad
b=-\Theta_{(\FF)},
\qquad
c=\widetilde\Sigma_{\FF}.
\end{equation}
and therefore the evolution of $\chi_{\FF}$ is controlled by the competition between nonlinear self-coupling, kinematical damping or amplification through $\Theta_{(\FF)}$, and the effective multi-scalar source $\widetilde\Sigma_{\FF}$.

The instantaneous critical lines are determined by
\begin{equation}
8\pi\chi_{\FF}^{\,2}
-\Theta_{(\FF)}\chi_{\FF}
+\widetilde\Sigma_{\FF}=0,
\label{eq:criticallineeq}
\end{equation}
namely
\begin{equation}
\chi_{\FF}^{(\pm)}
=
\frac{\Theta_{(\FF)}\pm\sqrt{\Theta_{(\FF)}^2-32\pi\widetilde\Sigma_{\FF}}}{16\pi},
\quad
\Theta_{(\FF)}^2\ge 32\pi\widetilde\Sigma_{\FF}.
\label{eq:criticalroots}
\end{equation}
A particularly transparent subcase is
\begin{equation}
\widetilde\Sigma_{\FF}=0
\qquad\Longleftrightarrow\qquad
\Box\FF=0,
\label{eq:criticalsubcase}
\end{equation}
for which
\begin{equation}
\dot\chi_{\FF}
=
\chi_{\FF}\left(8\pi\chi_{\FF}-\Theta_{(\FF)}\right).
\label{eq:KTFcriticalsimple}
\end{equation}

Since $\chi_{\FF}=K_{\FF}T_{\FF}\ge0$ on the physical Eckart branch, the sign of $\dot\chi_{\FF}$ has a direct thermodynamic meaning:
\begin{equation}
\dot\chi_{\FF}>0
\quad\Longrightarrow\quad
\text{heating of the coupling channel},
\end{equation}
whereas
\begin{equation}
\dot\chi_{\FF}<0
\quad\Longrightarrow\quad
\text{cooling of the coupling channel}.
\end{equation}

This interpretation is specific to the effective coupling variable.

\subsubsection{$\chi^{A}$ sector}

The variable $\chi_{\FF}$ governs only the projection of the thermal state along the coupling direction. To follow the full scalar thermal dynamics one must track the field-space vector $\chi^A$ itself. From Eq.~\eqref{eq:chiAdef_transport}, the quantities $\chi^A$ are directly proportional to the scalar velocity vector in field space and therefore retain the full directional information about the time evolution of the scalar sector. By contrast, Eq.~\eqref{eq:chiFfromchiA_transport} shows that $\chi_{\FF}$ is only the projection of this vector along the coupling direction $\FF_{,A}$. Hence $\chi_{\FF}$ diagnoses the activity of the coupling channel alone, whereas $\chi^A$ resolves which scalar directions remain dynamically active after that projection has been taken.

Differentiating Eq.~\eqref{eq:chiAdef_transport} gives
\begin{equation}
\dot{\chi}^{A}
=
-\frac{1}{8\pi \FF}\,\ddot{\phi}^{A}_{(\FF)}
+\frac{\dot{\FF}_{(\FF)}\dot{\phi}^{A}_{(\FF)}}{8\pi \FF^{2}}
=
-\frac{1}{8\pi \FF}\,\ddot{\phi}^{A}_{(\FF)}
+8\pi\,\chi_{\FF}\chi^{A}.
\label{eq:chidotA_step2}
\end{equation} 

When $\mathcal{B}_{AB}$ is nondegenerate, a transport equation for $\chi^A$ follows from the field-space covariant scalar equations \eqref{eq:scalarmulticov} by using Eq.~\eqref{eq:boxdecomp} and Eq.~\eqref{eq:graddecomp} that yields
\begin{align}
\ddot{\phi}^{A}_{(\FF)}
&=
-\Theta_{(\FF)}\dot{\phi}^{A}_{(\FF)}
+\D^2_{(\FF)}\phi^A
+a^a_{(\FF)}\D_a^{(\FF)}\phi^A
\nonumber\\
&
+\Gamma^{A}{}_{BC}(\mathcal{B})
\left[
-\dot{\phi}^{B}_{(\FF)}\dot{\phi}^{C}_{(\FF)}
+\D_a^{(\FF)}\phi^B\,\D^a_{(\FF)}\phi^C
\right]
\nonumber\\
&
+\frac12\,\mathcal{B}^{AB}\left(\FF_{,B}R-U_{,B}\right).
\label{eq:ddotphiA_Fframe}
\end{align}
and thus Eq.~\eqref{eq:chidotA_step2} can be recast as
\begin{align}
\dot{\chi}^{A}
&=
\left(8\pi\chi_{\FF}-\Theta_{(\FF)}\right)\chi^{A}
+8\pi \FF\,\Gamma^{A}{}_{BC}(\mathcal{B})\,\chi^{B}\chi^{C}
\nonumber\\
&-\frac{1}{8\pi \FF}\left[\D^2_{(\FF)}\phi^A
+a^a_{(\FF)}\D_a^{(\FF)}\phi^A \right.\nonumber \\
&\left.+\Gamma^{A}{}_{BC}(\mathcal{B})\,\D_a^{(\FF)}\phi^B\,\D^a_{(\FF)}\phi^C \right]
+\frac{1}{16\pi \FF}\left( U^{,A}-\FF^{,A}R \right),
\label{eq:chiAtransport}
\end{align}
with
\begin{equation}
U^{,A}\equiv \mathcal{B}^{AB}U_{,B},
\qquad
\FF^{,A}\equiv \mathcal{B}^{AB}\FF_{,B}.
\label{eq:raisedFU}
\end{equation}

It is useful to express this in field-space covariant form. Defining
\begin{equation}
\frac{\mathcal D\chi^{A}}{d\tau}
\equiv
\dot{\chi}^{A}
+\Gamma^{A}{}_{BC}(\mathcal{B})\,\dot{\phi}^{B}_{(\FF)}\chi^{C},
\label{eq:covderchiA}
\end{equation}
and using $\dot{\phi}^{B}_{(\FF)}=-8\pi\FF\chi^{B}$, one obtains
\begin{align}
\frac{\mathcal{D}\chi^{A}}{d\tau}
&=
\left(8\pi\chi_{\FF}-\Theta_{(\FF)}\right)\chi^{A}+\frac{1}{16\pi \FF}\left( U^{,A}-\FF^{,A}R \right)
\nonumber\\
&-\frac{1}{8\pi \FF}\left[\D^2_{(\FF)}\phi^A
+a^a_{(\FF)}\D_a^{(\FF)}\phi^A \right.\nonumber \\
&\left.+\Gamma^{A}{}_{BC}(\mathcal{B})\,\D_a^{(\FF)}\phi^B\,\D^a_{(\FF)}\phi^C \right],
\label{eq:chiAtransport_cov}
\end{align}

The previous equation governs the intrinsic evolution of the scalar thermal vector in field space. Each scalar direction is driven by four distinct contributions: a universal damping or amplification term, a nonlinear field-space coupling, a spatial-gradient sector, and an effective force term.

On the purely inertial branch,
\begin{equation}
\D_a^{(\FF)}\phi^A=0,
\label{eq:inertialbranchmulti_repeat}
\end{equation}
the transport equation simplifies to
\begin{equation}
\frac{\mathcal{D}\chi^{A}}{d\tau}
=
\left(8\pi\chi_{\FF}-\Theta_{(\FF)}\right)\chi^{A}
+\frac{1}{16\pi \FF}
\left(
U^{,A}-\FF^{,A}R
\right).
\label{eq:chiAtransport_inertial}
\end{equation}

\subsubsection{$\chi_A$ sector}

The lower-index quantity $\chi_A$ is the constitutive thermal covector selected by the kinetic matrix. It is the natural object in the heat flux, in the residual temperature-gradient sector, and in channel-resolved constitutive statements. If $\mathcal{B}_{AB}$ is nondegenerate, Eq.~\eqref{eq:chiAtransport_cov} can simply be lowered. More importantly, however, a transport equation for $\chi_A$ can be obtained directly from the lower-index scalar equations and therefore remains valid even when $\mathcal{B}_{AB}$ is degenerate.

Starting from Eq.~\eqref{eq:scalarmulti} we define
\begin{equation}
\Upsilon_A\equiv \mathcal{B}_{AB}\dot\phi^B_{(\FF)},
\qquad
\chi_A\equiv -\frac{\Upsilon_A}{8\pi\FF}.
\label{eq:UpsilonChiLower}
\end{equation}
Since
\begin{equation}
\dot\Upsilon_A
=
\mathcal B_{AB,C}\dot\phi^C_{(\FF)}\dot\phi^B_{(\FF)}
+
\mathcal{B}_{AB}\ddot\phi^B_{(\FF)},
\end{equation}
a direct computation yields
\begin{align}
\dot\Upsilon_A
&=
-\Theta_{(\FF)}\,\Upsilon_A
+
\mathcal{B}_{AB}
\left(
\D^2_{(\FF)}\phi^B
+
a^a_{(\FF)}\D_a^{(\FF)}\phi^B
\right)
\nonumber\\
&
+\frac12\,\mathcal B_{BC,A}\,
\dot\phi^B_{(\FF)}\dot\phi^C_{(\FF)}
\nonumber\\
&
+\frac12
\left(
2\mathcal B_{AB,C}-\mathcal B_{BC,A}
\right)
\D_a^{(\FF)}\phi^B\,\D^a_{(\FF)}\phi^C
\nonumber\\
&
+\frac12\left(\FF_{,A}R-U_{,A}\right).
\label{eq:UpsilonTransport}
\end{align}
Differentiating $\chi_A=-\Upsilon_A/(8\pi\FF)$ then gives
\begin{align}
\dot\chi_A
&=\left(8\pi\chi_{\FF}-\Theta_{(\FF)}\right)\chi_A\nonumber\\
&-\frac{1}{8\pi\FF}\,\mathcal{B}_{AB}\left(\D^2_{(\FF)}\phi^B+a^a_{(\FF)}\D_a^{(\FF)}\phi^B\right)\nonumber\\
&-\frac{1}{16\pi\FF}\left(\mathcal B_{BC,A}\,\dot\phi^B_{(\FF)}\dot\phi^C_{(\FF)}+\FF_{,A}R-U_{,A}\right)\nonumber\\
&-\frac{1}{16\pi\FF}\left(2\mathcal B_{AB,C}-\mathcal B_{BC,A}\right)\D_a^{(\FF)}\phi^B\,\D^a_{(\FF)}\phi^C.
\label{eq:chiAlowerTransport}
\end{align}

Equation~\eqref{eq:chiAlowerTransport} is the general transport law for the constitutive thermal covector. It remains meaningful even when $\mathcal{B}_{AB}$ is degenerate. This is precisely why $\chi_A$ is indispensable: it is the quantity that preserves direct constitutive meaning even in theories for which $\chi^A$ cannot be reconstructed uniquely from $\chi_A$.

\subsubsection{$W_a$ sector}

The residual temperature-gradient sector must also be evolved. Since $W_a^{(\FF)}$ is spatial, its natural evolution variable is the projected time derivative
\begin{equation}
\dot W_{\langle a\rangle}^{(\FF)}
\equiv
h_a{}^{b}\,u^c_{(\FF)}\nabla_c W_b^{(\FF)},
\label{eq:Wdotdef}
\end{equation}
with
\begin{equation}
V_{\langle a\rangle}\equiv h_a{}^{b}V_b.
\label{eq:singleindexproj}
\end{equation}

Because $W_a^{(\FF)}$ is built from the contraction of $\chi_A$ with the projected scalar gradients, it is not independent of the field-space thermal amplitudes. It is useful to introduce the field-space flux covector
\begin{equation}
\mathcal W_a^{(\FF)}
\equiv
\chi_A\,\D_a^{(\FF)}\phi^A,
\label{eq:mathcalWdef}
\end{equation}
so that
\begin{equation}
W_a^{(\FF)}
=
-\frac{\mathcal W_a^{(\FF)}}{\chi_{\FF}}.
\label{eq:WfrommathcalW}
\end{equation}
A direct application of Eq.~\eqref{eq:keyidentity2} gives, in the irrotational $\FF$-frame,
\begin{align}
\dot{\mathcal W}_{\langle a\rangle}^{(\FF)}
&=
\dot\chi_A\,\D_a^{(\FF)}\phi^A
\nonumber\\
&
+\chi_A
\left[
\D_a\dot\phi^A
-\left(\sigma_a{}^b+\frac13\Theta h_a{}^b\right)\D_b\phi^A
+\dot\phi^A a_a
\right]_{(\FF)}.
\label{eq:mathcalWtransport}
\end{align}
It is convenient to denote the right-hand side by
\begin{align}
\mathcal S_a^{(\FF)}&\equiv\dot\chi_A\,\D_a^{(\FF)}\phi^A\nonumber\\
&+\chi_A\left[\D_a\dot\phi^A-\left(\sigma_a{}^b+\frac13\Theta h_a{}^b\right)\D_b\phi^A+\dot\phi^A a_a\right]_{(\FF)},
\label{eq:SadefAnalysis}
\end{align}
so that
\begin{equation}
\dot{\mathcal W}_{\langle a\rangle}^{(\FF)}
=
\mathcal S_a^{(\FF)}.
\label{eq:mathcalWtransportCompact}
\end{equation}

Using Eq.~\eqref{eq:WfrommathcalW}, one then finds
\begin{equation}
\dot W_{\langle a\rangle}^{(\FF)}
=
-\frac{\dot\chi_{\FF}}{\chi_{\FF}}\,W_a^{(\FF)}
-\frac{1}{\chi_{\FF}}\,\mathcal S_a^{(\FF)}.
\label{eq:WtransportCompact}
\end{equation}

Both Eq.~\eqref{eq:mathcalWtransport} and Eq.~\eqref{eq:WtransportCompact} are transport laws for the residual temperature-gradient sector. The latter shows that the evolution of $W_a^{(\FF)}$ is controlled by two coupled mechanisms: a homogeneous rescaling governed by the coupling amplitude $\chi_{\FF}$ and an inhomogeneous source $\mathcal S_a^{(\FF)}$ built from the evolution of the constitutive thermal covector $\chi_A$ together with the projected scalar gradients. In this precise sense, $\chi_A$ and $W_a^{(\FF)}$ are not independent. The former selects which scalar directions contribute kinetically to the heat sector; the latter measures how that kinetic information survives as a genuine temperature-gradient contribution once it is contracted with the spatial scalar structure.  It is important, however, to distinguish between the weaker condition $\dot W_{\langle a\rangle}^{(\FF)}\to0$, which only implies asymptotic freezing, and the stronger statement $W_a^{(\FF)}\to0$, which implies the disappearance of the residual temperature-gradient sector. A clear example of this point is given by considering
\begin{equation}
\mathcal S_a^{(\FF)}\rightarrow 0.
\label{eq:Sadecay}
\end{equation}
Then Eq.~\eqref{eq:WtransportCompact} reduces asymptotically to
\begin{equation}
\dot W_{\langle a\rangle}^{(\FF)}
\simeq
-\frac{\dot\chi_{\FF}}{\chi_{\FF}}\,W_a^{(\FF)},
\label{eq:Whomogeneousasymp}
\end{equation}
so that the sign of $\dot \chi_{\FF}/\chi_{\FF}$ controls the local behavior of the homogeneous part:
\begin{equation}
\frac{\dot\chi_{\FF}}{\chi_{\FF}}>0
\quad\Longrightarrow\quad
\text{damping of }W_a^{(\FF)},
\label{eq:dampingcriterionW}
\end{equation}
whereas
\begin{equation}
\frac{\dot\chi_{\FF}}{\chi_{\FF}}<0
\quad\Longrightarrow\quad
\text{amplification of }W_a^{(\FF)},
\label{eq:amplificationcriterionW}
\end{equation}
and
\begin{equation}
\frac{\dot\chi_{\FF}}{\chi_{\FF}}\to 0
\quad\Longrightarrow\quad
W_a^{(\FF)} \ \text{sets into a constant value}.
\end{equation}

\subsection{Thermal diagnostics} \label{subsec:diagnostics}

The transport system above shows that three distinct notions of thermal evolution must be distinguished.

The first is the evolution of the \emph{coupling channel}. This is diagnosed by $\chi_{\FF}$ and its transport equation~\eqref{eq:chiFtransportSigma}. Since $\chi_{\FF}=K_{\FF}T_{\FF}\ge0$ on the physical Eckart branch, the sign of $\dot\chi_{\FF}$ has an immediate thermodynamic interpretation: it tells whether the coupling channel heats up or cools down. This is the genuine multi-scalar extension of the one-field $KT$ criterion. Its meaning, however, is restricted: it probes only the effective coupling sector and says nothing by itself about scalar directions orthogonal to $\FF_{,A}$ in field space.

The second notion concerns the evolution of the full time-like multi-scalar thermal state. In general, this cannot be inferred from $\chi_{\FF}$ alone, because $\chi_{\FF}$ probes only the projection $\FF_{,A}\chi^A$ of the field-space thermal vector along the coupling direction. To determine whether the complete scalar velocity sector is relaxing toward equilibrium, characterized by $\chi^A\to0$ for all $A$, or instead moving away from it, one needs a scalar quantity that measures the magnitude of the full field-space thermal vector. The natural candidate is
\begin{equation}
\mathfrak D_\chi \equiv \chi_A\chi^A=\mathcal{B}_{AB}\chi^A\chi^B.
\label{eq:Dchi_def}
\end{equation}
that we will call the time-like thermal magnitude. This quantity is not an effective temperature. Rather, it is the scalar obtained by contracting the field-space thermal vector with the kinetic matrix of the theory itself. In this sense it is the canonical scalar built from the time-like thermal amplitudes whenever $\mathcal{B}_{AB}$ is invertible. The need for $\mathfrak D_\chi$ arises because $\chi_A$ and $\chi^A$ are generically signed field-space components, so their individual values do not by themselves measure the magnitude of the full time-like thermal state. Therefore the sign of $\dot\chi^A$ or $\dot\chi_A$ alone does not determine whether the full scalar thermal state is moving toward or away from equilibrium.

The third and final notion is the evolution of the \emph{spatial} scalar structure. To monitor this part of the dynamics it is natural to introduce, in complete analogy with $\mathfrak D_\chi$, the scalar
\begin{equation}
\mathfrak D_{\rm grad}
\equiv\mathcal{B}_{AB}\,\D_a^{(\FF)}\phi^A\,\D^a_{(\FF)}\phi^B,
\label{eq:Dgrad_def}
\end{equation}
that we will refer to as the spatial gradient magnitude. This quantity is the canonical scalar built from the spatial multi-scalar sector using the field-space kinetic matrix and the positive-definite metric induced on the local rest spaces. Thus $\mathfrak D_{\rm grad}$ is the natural scalar measure of the magnitude of the residual spatial scalar configuration, just as $\mathfrak D_\chi$ measures the magnitude of the time-like thermal one.

The interpretation of $\mathfrak D_\chi$ and $\mathfrak D_{\rm grad}$ depends crucially on the properties of $\mathcal{B}_{AB}$. If $\mathcal{B}_{AB}$ is merely nondegenerate, then $\mathfrak D_\chi$ and $\mathfrak D_{\rm grad}$ are well-defined scalars, but it need not be positive. In that case it is still a useful invariant contraction, but it should not yet be interpreted as a norm. To obtain that stronger interpretation one must further require that the kinetic matrix be positive definite on the physical scalar sector, namely
\begin{equation}
\mathcal{B}_{AB}X^A X^B>0
\quad
\text{for every field-space vector }X^A.
\label{eq:B_positive_definite}
\end{equation}
Under this stronger assumption one has
\begin{equation}
\mathfrak D_\chi\ge0,
\qquad
\mathfrak D_\chi=0
\quad\Longleftrightarrow\quad
\chi^A=0
\qquad
\forall A,
\label{eq:Dchi_positive}
\end{equation}
and
\begin{equation}
\mathfrak D_{\rm grad}\ge0,
\qquad
\mathfrak D_{\rm grad}=0
\quad\Longleftrightarrow\quad
\D_a^{(\FF)}\phi^A=0
\qquad
\forall A.
\label{eq:Dgrad_positive}
\end{equation}
so that $\mathfrak D_\chi$ becomes a genuine norm-like measure of the magnitude of the full time-like thermal vector and $\mathfrak D_{\rm grad}$ becomes a genuine norm-like measure of the magnitude of the spatial scalar sector.

The assumption of $\mathcal{B}_{AB}$ being positive definite is physically natural but not automatic. In healthy multi-scalar-tensor theories one commonly requires the scalar kinetic sector to be positive definite, at least on the physical branch, in order to avoid ghostlike directions and to ensure standard-sign kinetic energy. In that sense the condition~\eqref{eq:B_positive_definite} is not exotic, but it is nevertheless an additional assumption rather than a universal identity of the theory. The appropriate hierarchy is therefore the following: nondegeneracy of $\mathcal{B}_{AB}$ is sufficient to define $\mathfrak D_\chi$ and $\mathfrak D_{\rm grad}$ canonically; positive definiteness is additionally required if one wishes to interpret them as genuine positive measures of the full time-like multi-scalar thermal state and of the spatial scalar-gradient thermal sector.

Before advancing into the analysis, it is useful to collect the main coupling-frame and thermodynamic variables that have appeared so far. Table~\ref{tab:thermal_dictionary} summarizes their definitions and physical meaning. The table is meant only as a compact guide to the notation used in the transport and diagnostic analysis; the standard $1+3$ kinematical quantities were defined earlier and are not repeated.
\begin{table*}[t!]
\caption{Compact dictionary of the coupling-frame and first-order thermodynamic variables.}
\label{tab:thermal_dictionary}
\centering
\renewcommand{\arraystretch}{1.25}
\scriptsize
\begin{tabular}{lll}
\hline\hline
{Symbol} & {Definition} & {Physical meaning} \\
\hline

$\mathcal B_{AB}$ &
Kinetic matrix of the scalar sector &
Field-space metric only if nondegenerate; norm-defining only if positive definite \\

$u_a^{(\FF)}$ &
{$u_a^{(\FF)}=\nabla_a\FF/\sqrt{-\nabla_b\FF\nabla^b\FF}$} &
{Frame comoving with the effective coupling, defined when $\nabla_a\FF$ is timelike} \\

{$\chi_{\FF}$} &
{$\chi_{\FF}=K_{\FF}T_{\FF}=-\dot{\FF}_{(\FF)}/(8\pi\FF)$} &
{Inertial thermal variable associated with the coupling channel} \\

{$W_a^{(\FF)}$} &
{$W_a^{(\FF)}=-(\dot{\FF})^{-1}\mathcal B_{AB}\dot{\phi}^A\D_a^{(\FF)}\phi^B$} &
{Residual spatial multi-scalar sector in the coupling frame} \\

{$T_{\FF}$} &
{$\D_a^{(\FF)}\ln T_{\FF}=W_a^{(\FF)}$, when integrable} &
{Effective geometric temperature in the coupling frame} \\

{$K_{\FF}$} &
{$K_{\FF}=\chi_{\FF}/T_{\FF}$, once $T_{\FF}$ is fixed} &
{Effective first-order thermal conductivity} \\

{$\chi^A$} &
{$\chi^A=-\dot{\phi}^A_{(\FF)}/(8\pi\FF)$} &
{Field-space thermal vector encoding the time-like scalar dynamics} \\

{$\chi_A$} &
{$\chi_A=\mathcal B_{AB}\chi^B$} &
{Kinetic-weighted thermal covector} \\

{$\mathfrak D_\chi$} &
{$\mathfrak D_\chi=\chi_A\chi^A$} &
{Time-like scalar diagnostic; nonnegative only if $\mathcal B_{AB}$ is positive definite} \\

{$\mathfrak D_{\rm grad}$} &
{$\mathfrak D_{\rm grad}=\mathcal B_{AB}\D_a^{(\FF)}\phi^A\D^a_{(\FF)}\phi^B$} &
{Spatial scalar diagnostic; nonnegative only if $\mathcal B_{AB}$ is positive definite} \\

\hline\hline
\end{tabular}
\end{table*}

\subsubsection{$\mathfrak D_\chi$ analysis}
As for $\chi_{\FF}$, we now derive the evolution equation for $\mathfrak D_\chi$. Taking the proper-time derivative along the $\FF$-flow yields
\begin{equation}
\dot{\mathfrak D}_\chi=\frac{d}{d\tau}(\chi_A\chi^A)=2\chi_A\,\frac{\mathcal D\chi^A}{d\tau}.
\label{eq:Dchi_dot}
\end{equation}
that by substituting Eq.~\eqref{eq:chiAtransport_cov} into Eq.~\eqref{eq:Dchi_dot}, can be written as
\begin{align}
\dot{\mathfrak D}_\chi
&=
2\left(8\pi\chi_{\FF}-\Theta_{(\FF)}\right)\mathfrak D_\chi+\frac{1}{8\pi\FF}\,\chi_A\left(U^{,A}-\FF^{,A}R\right)
\nonumber\\
&-\frac{1}{4\pi\FF}\,\chi_A
\left[
\D^2_{(\FF)}\phi^A
+a^a_{(\FF)}\D_a^{(\FF)}\phi^A \right. \nonumber\\
&\left.+\Gamma^{A}{}_{BC}(\mathcal B)\,
\D_a^{(\FF)}\phi^B\D^a_{(\FF)}\phi^C
\right].
\label{eq:Dchi_transport_full}
\end{align}
It is useful to collect the nonuniversal terms into the effective source
\begin{align}
\widetilde\Sigma_\chi
&\equiv
-\frac{1}{4\pi\FF}\,
\chi_A
\left[
\D^2_{(\FF)}\phi^A
+a^a_{(\FF)}\D_a^{(\FF)}\phi^A\right. 
\nonumber\\
&\left.+\Gamma^{A}{}_{BC}(\mathcal B)\,
\D_a^{(\FF)}\phi^B\D^a_{(\FF)}\phi^C
\right]+\frac{1}{8\pi\FF}\,
\chi_A\left(U^{,A}-\FF^{,A}R\right),
\label{eq:SigmaChi_def}
\end{align}
so that the transport law becomes
\begin{equation}
\dot{\mathfrak D}_\chi
=
2\left(8\pi\chi_{\FF}-\Theta_{(\FF)}\right)\mathfrak D_\chi
+\widetilde\Sigma_\chi.
\label{eq:Dchi_transport_sigma}
\end{equation}

Equation~\eqref{eq:Dchi_transport_sigma} is the natural analogue, for the full time-like multi-scalar sector, of the coupling-channel transport law~\eqref{eq:chiFtransportSigma}: when $\dot{\mathfrak D}_\chi<0$, the magnitude of the time-like multi-scalar thermal vector decreases; when $\dot{\mathfrak D}_\chi>0$, the full time-like sector increases. There is, however, an important structural difference. The equation for $\chi_{\FF}$ is closed once $\widetilde\Sigma_{\FF}$ is specified, whereas Eq.~\eqref{eq:Dchi_transport_sigma} is not in general a closed scalar equation for $\mathfrak D_\chi$ alone, because the source $\widetilde\Sigma_\chi$ depends on the full field-space configuration, including the direction of $\chi^A$ and the projected scalar gradients. Accordingly, the critical set defined by $\dot{\mathfrak D}_\chi=0$ is not, in general, a critical \emph{line} but a critical hypersurface in the full multi-scalar state space,
\begin{equation}
2\left(8\pi\chi_{\FF}-\Theta_{(\FF)}\right)\mathfrak D_\chi
+\widetilde\Sigma_\chi
=0.
\label{eq:Dchi_critical_hypersurface}
\end{equation}

Even so, the universal part of Eq.~\eqref{eq:Dchi_transport_sigma} already has a clear physical interpretation. The coefficient
\begin{equation}
2\left(8\pi\chi_{\FF}-\Theta_{(\FF)}\right)
\label{eq:Dchi_universal_coeff}
\end{equation}
controls the homogeneous growth or decay of the full time-like thermal magnitude. In an expanding coupling congruence, $\Theta_{(\FF)}>0$, the expansion tends to damp $\mathfrak D_\chi$, whereas a sufficiently large coupling-channel activity $\chi_{\FF}$ tends to amplify it. Thus, disregarding the source for the moment, one has the threshold
\begin{equation}
8\pi\chi_{\FF}=\Theta_{(\FF)},
\label{eq:Dchi_threshold}
\end{equation}
which separates homogeneous damping from homogeneous amplification.

A particularly transparent subcase is the source-free branch
\begin{equation}
\widetilde\Sigma_\chi=0,
\label{eq:Dchi_sourcefree}
\end{equation}
for which Eq.~\eqref{eq:Dchi_transport_sigma} reduces to
\begin{equation}
\dot{\mathfrak D}_\chi
=
2\mathfrak D_\chi
\left(8\pi\chi_{\FF}-\Theta_{(\FF)}\right),
\label{eq:Dchi_transport_simple}
\end{equation}
showing that evolution of the time-like scalar sector depends on the coupling dynamics through $\chi_\FF$.

For an expanding coupling congruence, $\Theta_{(\FF)}>0$, one obtains
\begin{equation}
0<\chi_{\FF}<\frac{\Theta_{(\FF)}}{8\pi}
\quad\Longrightarrow\quad
\dot{\mathfrak D}_\chi<0,
\label{eq:Dchi_cooling}
\end{equation}
so that the full time-like multi-scalar thermal sector is driven toward $\chi^A=0$, cooling, whereas
\begin{equation}
\chi_{\FF}>\frac{\Theta_{(\FF)}}{8\pi}
\quad\Longrightarrow\quad
\dot{\mathfrak D}_\chi>0,
\label{eq:Dchi_heating}
\end{equation}
so that the full time-like multi-scalar thermal sector moves away from $\chi^A=0$, heating. For a contracting coupling congruence, $\Theta_{(\FF)}<0$, one has
\begin{equation}
\dot{\mathfrak D}_\chi>0,
\label{eq:Dchi_contracting}
\end{equation}
so contraction universally amplifies the time-like multi-scalar thermal state on this branch. On the purely inertial branch, $\D_a^{(\FF)}\phi^A=0$, Eq.~\eqref{eq:Dchi_transport_full} simplifies to
\begin{equation}
\dot{\mathfrak D}_\chi
=
2\left(8\pi\chi_{\FF}-\Theta_{(\FF)}\right)\mathfrak D_\chi
+\frac{1}{8\pi\FF}\,\chi_A\left(U^{,A}-\FF^{,A}R\right).
\label{eq:Dchi_transport_inertial}
\end{equation}
Hence, when the residual temperature-gradient sector has disappeared, the full time-like multi-scalar state still evolves under the combined action of the universal damping/amplification term and the projection of the scalar-force sector along the thermal vector. If, in addition,
\begin{equation}
\chi_A\left(U^{,A}-\FF^{,A}R\right)=0,
\label{eq:Dchi_inertial_sourcefree}
\end{equation}
then Eq.~\eqref{eq:Dchi_transport_inertial} reduces again to the source-free form~\eqref{eq:Dchi_transport_simple}, and the sign conditions~\eqref{eq:Dchi_cooling}-\eqref{eq:Dchi_contracting} hold exactly.

Finally, it is useful to compare $\mathfrak D_\chi$ with the coupling projection itself. When $\mathcal{B}_{AB}$ is nondegenerate, one may decompose the thermal vector into a component parallel to $\FF_{,A}$ and a component orthogonal to it in field space. Writing
\begin{equation}
Q\equiv \FF_{,A}\FF^{,A} \neq0,
\quad
\chi^A=
\frac{\chi_{\FF}}{Q}\,\FF^{,A}
+\chi_\perp^A,
\quad
\FF_{,A}\chi_\perp^A=0,
\label{eq:chi_decomp_parallel}
\end{equation}
one finds
\begin{equation}
\mathfrak D_\chi
=
\chi_A\chi^A
=
\frac{\chi_{\FF}^2}{Q}
+\chi_{\perp A}\chi_\perp^A.
\label{eq:Dchi_decomp}
\end{equation}

This decomposition makes explicit why $\chi_{\FF}$ is not sufficient to diagnose the full approach to GR: it probes only the first term, whereas $\mathfrak D_\chi$ also captures the scalar directions orthogonal to the coupling channel. Equation~\eqref{eq:Dchi_transport_sigma} therefore provides the natural scalar criterion for approach to, or departure from, full multi-scalar equilibrium in the time-like sector. Unlike $\dot\chi_{\FF}$, which probes only the coupling channel, the sign of $\dot{\mathfrak D}_\chi$ determines whether the magnitude of the full scalar thermal vector is shrinking or growing. In this precise sense, $\chi_{\FF}$ diagnoses heating or cooling of the coupling projection, whereas $\mathfrak D_\chi$ diagnoses relaxation or departure of the full time-like multi-scalar state.

Furthermore, as previously stressed, the interpretation and use of the scalar $\mathfrak{D}_{\chi}$, and by extension of its evolution equation, are tied to the properties of the kinetic matrix $\mathcal{B}_{AB}$. If $\mathcal{B}_{AB}$ is invertible but indefinite, then $\mathfrak D_\chi$ remains canonical but its evolution equation does not have the ability to measure if the time-like multi-scalar thermal vector is increasing or decreasing. If $\mathcal{B}_{AB}$ is degenerate, the situation is more complex: there is then no canonical scalar of the form $\chi_A\chi^A$ with norm-like meaning, because $\mathcal{B}_{AB}$ no longer defines a genuine metric on field space and there is, in general, no canonical closed transport equation for $\mathfrak D_\chi$. In particular, $\chi_A\chi^A$ may vanish while some components of $\chi^A$ remain nonzero along kernel directions of the kinetic matrix. Accordingly, in the degenerate case no unique model-independent scalar exists that faithfully measures the full time-like evolution. The complete analysis must then be carried out directly in terms of the vector $\chi^A$ itself, or else using an additional positive-definite field-space structure supplied by the specific model. This does not represent a limitation of the thermodynamic interpretation itself, but only of the possibility of encoding the full time-like multi-scalar state in a unique model-independent scalar.

\subsubsection{$\mathfrak D_{\rm grad}$ analysis}

We will now focus on the spatial sector. Once more, the proper-time derivative of $\mathfrak D_{\rm grad}$ is obtained by differentiating its definition along the $\FF$-flow. Since $\D_a^{(\FF)}\phi^A$ is purely spatial, the derivative of the projector does not contribute after the full contraction, and one finds
\begin{equation}
\dot{\mathfrak D}_{\rm grad}
=
\dot{\mathcal B}_{AB}\,\D_a^{(\FF)}\phi^A\,\D^a_{(\FF)}\phi^B
+
2\mathcal{B}_{AB}\,\D^a_{(\FF)}\phi^A\,
\dot{\D}_{\langle a\rangle}^{(\FF)}\phi^B.
\label{eq:Dgrad_derivation_start}
\end{equation}

Again, the characteristics of $\mathcal{B}_{AB}$ determine the use and interpretation of $\mathfrak D_{\text{grad}}$ and its evolution equation. Moreover, the projected-gradient evolution is obtained by applying Eq.~\eqref{eq:keyidentity2} to $\varphi=\phi^A$ and then using Eq.~\eqref{eq:chiAdef_transport} together with $\D_a^{(\FF)}\FF=0$. In the irrotational $\FF$-frame this gives
\begin{align}
\dot{\D}_{\langle a\rangle}^{(\FF)}\phi^A&=-8\pi\FF\,\D_a^{(\FF)}\chi^A
-\left(\sigma_a{}^b+\frac13\Theta h_a{}^b\right)_{(\FF)} \D_b^{(\FF)}\phi^A \nonumber\\
&-8\pi\FF\,\chi^A a_a^{(\FF)}.
\label{eq:gradphi_transport}
\end{align}
and by substituting it into Eq.~\eqref{eq:Dgrad_derivation_start} yields
\begin{align}
\dot{\mathfrak D}_{\rm grad}
&=\dot{\mathcal B}_{AB}\,\D_a^{(\FF)}\phi^A\,\D^a_{(\FF)}\phi^B
-16\pi\FF\,\mathcal{B}_{AB}\,\D^a_{(\FF)}\phi^A\,\D_a^{(\FF)}\chi^B
\nonumber\\
&-2\left(\sigma^{ab}+\frac13\Theta h^{ab}\right)_{(\FF)}
\mathcal{B}_{AB}\,\D_a^{(\FF)}\phi^A\,\D_b^{(\FF)}\phi^B
\nonumber\\
&-16\pi\FF\,a^a_{(\FF)}\,\chi_A\,\D_a^{(\FF)}\phi^A,
\label{eq:Dgrad_derivation_intermediate}
\end{align}
where $\chi_A\equiv \mathcal{B}_{AB}\chi^B$. Since $\mathcal{B}_{AB}$ depends only on the scalar fields one has
\begin{equation}
\dot{\mathcal B}_{AB}
=
\mathcal B_{AB,C}\,\dot\phi^C_{(\FF)}
=
-8\pi\FF\,\mathcal B_{AB,C}\,\chi^C.
\label{eq:Bdot_Fframe}
\end{equation}

Furthermore, it is convenient to use the field-space covariant projected derivative
\begin{equation}
\mathcal D_a^{(\FF)}\chi^A
\equiv
\D_a^{(\FF)}\chi^A
+
\Gamma^{A}{}_{BC}(\mathcal B)\,
\D_a^{(\FF)}\phi^B\,\chi^C,
\label{eq:spatial_cov_derivative_chi}
\end{equation}
and by using the metric compatibility of the Levi-Civita connection of $\mathcal{B}_{AB}$,
\begin{equation}
\mathcal B_{AB,C}
=
\Gamma^{D}{}_{AC}(\mathcal B)\,\mathcal B_{DB}
+
\Gamma^{D}{}_{BC}(\mathcal B)\,\mathcal B_{AD},
\label{eq:B_metric_compatibility}
\end{equation}
the first two terms in Eq.~\eqref{eq:Dgrad_derivation_intermediate} combine into
\begin{align}
&\dot{\mathcal B}_{AB}\,\D_a^{(\FF)}\phi^A\,\D^a_{(\FF)}\phi^B
-16\pi\FF\,\mathcal{B}_{AB}\,\D^a_{(\FF)}\phi^A\,\D_a^{(\FF)}\chi^B
\nonumber\\
&
=-16\pi\FF\,\mathcal{B}_{AB}\,\D_{(\FF)}^{a}\phi^A\,\mathcal D_a^{(\FF)}\chi^B.
\label{eq:Bdot_combine}
\end{align}

Substituting Eq.~\eqref{eq:Bdot_combine} back into Eq.~\eqref{eq:Dgrad_derivation_intermediate} gives 
\begin{align}
\dot{\mathfrak D}_{\rm grad}
&=
-16\pi\FF\,
\mathcal{B}_{AB}\,
\D_{(\FF)}^{a}\phi^A\,
\mathcal D_a^{(\FF)}\chi^B
\nonumber\\
&
-2\left(\sigma^{ab}+\frac13\Theta h^{ab}\right)_{(\FF)}
\mathcal{B}_{AB}\,
\D_a^{(\FF)}\phi^A\,
\D_b^{(\FF)}\phi^B
\nonumber\\
&
-16\pi\FF\,a^a_{(\FF)}\,\chi_A\,\D_a^{(\FF)}\phi^A,
\label{eq:Dgrad_transport_full}
\end{align}
where $\chi_A\equiv \mathcal{B}_{AB}\chi^B$. To better interpret the previous equation, it is useful to separate the trace-free part of the spatial gradient tensor by defining
\begin{equation}
\mathfrak S^{(\rm grad)}_{ab}
\equiv
\mathcal{B}_{AB}\,
\D_{\langle a}^{(\FF)}\phi^A\,
\D_{b\rangle}^{(\FF)}\phi^B,
\label{eq:Sgrad_def}
\end{equation}
using
\begin{equation}
\mathcal{B}_{AB}\,
\D_a^{(\FF)}\phi^A\,
\D_b^{(\FF)}\phi^B
=
\mathfrak S^{(\rm grad)}_{ab}
+\frac13\,h_{ab}^{(\FF)}\,\mathfrak D_{\rm grad},
\label{eq:Dgrad_tensor_split}
\end{equation}
equation~\eqref{eq:Dgrad_transport_full} can be written as
\begin{equation}
\dot{\mathfrak D}_{\rm grad}
=
-\frac23\,\Theta_{(\FF)}\,\mathfrak D_{\rm grad}
+\widetilde\Sigma_{\rm grad},
\label{eq:Dgrad_transport_sigma}
\end{equation}
with effective source
\begin{align}
\widetilde\Sigma_{\rm grad}
&\equiv
-2\,\sigma^{ab}_{(\FF)}\,\mathfrak S^{(\rm grad)}_{ab}-16\pi\FF\,
\mathcal{B}_{AB}\,
\D_{(\FF)}^{a}\phi^A\,
\mathcal D_a^{(\FF)}\chi^B
\nonumber\\
&
-16\pi\FF\,a^a_{(\FF)}\,\chi_A\,\D_a^{(\FF)}\phi^A.
\label{eq:SigmaGrad_def}
\end{align}

Equation~\eqref{eq:Dgrad_transport_sigma} is the spatial counterpart of Eq.~\eqref{eq:Dchi_transport_sigma}. Its interpretation is immediate. The universal term
\begin{equation}
-\frac23\,\Theta_{(\FF)}\,\mathfrak D_{\rm grad},
\label{eq:Dgrad_universal_term}
\end{equation}
describes the homogeneous dilution or amplification of the scalar-gradient sector by the expansion of the $\FF$-congruence, while $\widetilde\Sigma_{\rm grad}$ collects the genuinely nontrivial sources: anisotropic distortion through the contraction $\sigma^{ab}\mathfrak S^{(\rm grad)}_{ab}$, spatial inhomogeneity of the thermal vector through $\mathcal D_a^{(\FF)}\chi^A$, and acceleration-induced regeneration through $a^a_{(\FF)}\chi_A\D_a^{(\FF)}\phi^A$.

Just as in the time-like sector, the transport law for $\mathfrak D_{\rm grad}$ is not, in general, a closed scalar equation for $\mathfrak D_{\rm grad}$ alone. The source $\widetilde\Sigma_{\rm grad}$ depends on the full spatial configuration, including the anisotropic distribution of gradients encoded in $\mathfrak S^{(\rm grad)}_{ab}$ and the projected thermal derivatives $\mathcal D_a^{(\FF)}\chi^A$. Accordingly, the critical set for the spatial sector is not generically a line but a hypersurface in the full multi-scalar state space:
\begin{equation}
-\frac23\,\Theta_{(\FF)}\,\mathfrak D_{\rm grad}
+\widetilde\Sigma_{\rm grad}
=0.
\label{eq:Dgrad_critical_hypersurface}
\end{equation}

Nevertheless, the homogeneous part of Eq.~\eqref{eq:Dgrad_transport_sigma} already carries information about the dynamical evolution of $\mathfrak D_{\rm grad}$. In an expanding coupling congruence, $\Theta_{(\FF)}>0$, the expansion tends to damp the spatial scalar structure; in a contracting congruence, $\Theta_{(\FF)}<0$, it tends instead to amplify it. A particularly transparent subcase is the source-free branch
\begin{equation}
\widetilde\Sigma_{\rm grad}=0,
\label{eq:Dgrad_sourcefree}
\end{equation}
for which Eq.~\eqref{eq:Dgrad_transport_sigma} reduces to
\begin{equation}
\dot{\mathfrak D}_{\rm grad}
=
-\frac23\,\Theta_{(\FF)}\,\mathfrak D_{\rm grad}.
\label{eq:Dgrad_transport_simple}
\end{equation}
Hence, if $\mathfrak D_{\rm grad}>0$, one has
\begin{equation}
\Theta_{(\FF)}>0
\quad\Longrightarrow\quad
\dot{\mathfrak D}_{\rm grad}<0,
\label{eq:Dgrad_decay_expanding}
\end{equation}
so that the spatial scalar gradients are diluted along the flow, whereas
\begin{equation}
\Theta_{(\FF)}<0
\quad\Longrightarrow\quad
\dot{\mathfrak D}_{\rm grad}>0.
\label{eq:Dgrad_growth_contracting}
\end{equation}
Thus expansion drives the spatial multi-scalar sector toward homogeneity on this branch, while contraction amplifies it.

Equation~\eqref{eq:Dgrad_transport_sigma} therefore provides the natural scalar diagnostic for the spatial part of the GR-attractor problem. In precise analogy with the time-like sector, $\dot{\mathfrak D}_{\rm grad}<0$ indicates that the magnitude of the residual projected scalar structure is decaying, whereas $\dot{\mathfrak D}_{\rm grad}>0$ indicates that it is being regenerated or amplified. The crucial difference with $W_a^{(\FF)}$ is that $\mathfrak D_{\rm grad}$ captures the full projected scalar-gradient sector, while $W_a^{(\FF)}$ probes only the particular contraction selected by the constitutive thermal covector. For this reason $\mathfrak D_{\rm grad}$, rather than $W_a^{(\FF)}$, is the appropriate scalar quantity with which to analyze the spatial approach to the GR sector.

Taken together, $\chi_{\FF}$, $\mathfrak D_\chi$, and $\mathfrak D_{\rm grad}$ provide a natural hierarchy of diagnostics for the multi-scalar relaxation problem: $\chi_{\FF}$ tracks the coupling channel, $\mathfrak D_\chi$ tracks the full time-like multi-scalar thermal state, and $\mathfrak D_{\rm grad}$ tracks the full spatial scalar structure. The GR-like endpoint requires the joint decay of the last two, together with the induced vanishing of the coupling projection.

\subsection{Approach to GR criterion}
\label{subsec:GRcriterion}

We are now in a position to formulate the multi-scalar criterion for relaxation toward GR. In the recent one-field thermal literature, the approach to GR is tied to the vanishing of the single inertial thermal variable associated with the nonminimal coupling, which is equivalent to freezing the scalar degree of freedom~\cite{Faraoni:2025alq}. In the present multi-scalar theory, the closest analogue is the freezing of the \emph{coupling channel},
\begin{equation}
\nabla_a\FF=0,
\label{eq:couplingeq}
\end{equation}
and equivalently, in the $\FF$-comoving frame,
\begin{equation}
\chi_{\FF}=-\frac{\dot{\FF}_{(\FF)}}{8\pi\FF}\to0.
\label{eq:chiFcouplingeq}
\end{equation}

Equation~\eqref{eq:couplingeq} is therefore the natural multi-scalar generalization of the constant-coupling condition familiar from the one-field case. Its physical meaning is, however, weaker than full scalar relaxation. It freezes the effective coupling $\FF(\phi^A)$, but it does not in general freeze the entire scalar sector. This is precisely where the multi-scalar theory departs from the one-field picture: the coupling projection can relax even while scalar directions orthogonal to $\FF_{,A}$ in field space remain dynamically and thermodynamically active.

For this reason, a stronger notion of equilibrium is required. From the Einstein-like metric equations~\eqref{eq:einsteinlike_multi}, the natural definition of \emph{full multi-scalar equilibrium} is the constant-scalar sector
\begin{equation}
\nabla_a\phi^A=0
\qquad
\text{for all }A,
\label{eq:fullmultieq1}
\end{equation}
and since $\FF=\FF(\phi^A)$, this immediately implies $\nabla_a \FF=0$. Thus full multi-scalar equilibrium is a strictly stronger condition than coupling equilibrium. It is the genuine multi-scalar analogue of the GR equilibrium state: not only the coupling channel, but all independent scalar directions are frozen. The corresponding equilibrium points are not fixed by the coupling function alone. Indeed, setting $\nabla_a\phi^A=0$ in the scalar-field equations~\eqref{eq:scalarmulticov} gives
\begin{equation}
\FF_{,A}R_*-U_{,A}=0,
\label{eq:backgroundpointa_multi}
\end{equation}
or equivalently
\begin{equation}
U_{,A}(\phi_*)=\FF_{,A}(\phi_*)\,R_*.
\label{eq:backgroundpointa_multi2}
\end{equation}
Using the trace of the metric field equation then yields
\begin{equation}
2U_*-\FF_*R_*=8\pi T^{(m)}_*.
\label{eq:backgroundpointb_multi}
\end{equation}
Hence the equilibrium points are determined jointly by the constant-scalar field equations and the trace condition.

The GR-like endpoint of the theory is the constant-scalar sector
\begin{equation}
\nabla_a\phi^A\to0
\qquad
\forall A,
\label{eq:fullmultieqsplit0}
\end{equation}
which, in the $\FF$-comoving frame, is equivalent to
\begin{equation}
\dot{\phi}^A_{(\FF)}\to0,
\qquad
\D_a^{(\FF)}\phi^A\to0
\qquad
\forall A.
\label{eq:fullmultieqsplit}
\end{equation}
Thus the approach to the GR sector has two logically distinct components: a time-like one, corresponding to the freezing of the scalar velocities, and a spatial one, corresponding to the disappearance of the residual scalar gradients.

The time-like part is encoded by the field-space thermal vector $\chi^A$, so that
\begin{equation}
\chi^A\to0
\qquad\Longleftrightarrow\qquad
\dot{\phi}^A_{(\FF)}\to0
\qquad
\forall A.
\label{eq:chiAmeaning_equilibrium2}
\end{equation}
When $\mathcal{B}_{AB}$ is nondegenerate and positive definite on the physical scalar sector, this part may be summarized by the scalar $\mathfrak D_\chi$, with the transport law~\eqref{eq:Dchi_transport_sigma} providing the natural dynamical equation governing the time-like approach to the GR sector. More precisely, this equation determines whether the magnitude of the full scalar velocity vector in field space is decaying or growing along the $\FF$-flow. In configurations satisfying $\D_a^{(\FF)}\phi^A=0$, Eq.~\eqref{eq:Dchi_transport_inertial} decides whether the theory is cooling toward, or heating away from, time-like multi-scalar equilibrium.

The spatial part is encoded by the projected gradients themselves. When $\mathcal{B}_{AB}$ is positive definite, it is naturally summarized by $\mathfrak D_{\rm grad}$, so that
\begin{equation}
\mathfrak D_{\rm grad}\to0
\qquad\Longleftrightarrow\qquad
\D_a^{(\FF)}\phi^A\to0
\qquad
\forall A.
\label{eq:Dgrad_meaning_equilibrium}
\end{equation}
Thus, while $\dot\chi_{\FF}$ diagnoses heating or cooling of the coupling channel alone, $\dot{\mathfrak D}_\chi$ supplies the corresponding dynamical analysis for the full time-like multi-scalar sector, and $\dot{\mathfrak D}_{\rm grad}$ provides the complementary dynamical diagnostic for the spatial scalar sector. In this sense the multi-scalar approach to GR is governed not by a single thermal scalar, but by a coupled time-like and spatial relaxation problem.

The multi-scalar attractor criterion may therefore be stated as follows.

\medskip
\noindent
\textbf{Multi-scalar GR attractor criterion.}
A solution approaches the GR sector if and only if
\begin{equation}
\chi^A\to0,
\qquad
\D_a^{(\FF)}\phi^A\to0
\qquad
\forall A.
\label{eq:GRcriterion_main}
\end{equation}
If $\mathcal{B}_{AB}$ is nondegenerate, this is equivalently
\begin{equation}
\chi_A\to0,
\qquad
\D_a^{(\FF)}\phi^A\to0
\qquad
\forall A,
\label{eq:GRcriterion_main_lower}
\end{equation}
and, if moreover $\mathcal{B}_{AB}$ is positive definite, the time-like and spatial parts of the criterion may be summarized as
\begin{equation}
\mathfrak D_\chi\to0,
\qquad
\mathfrak D_{\rm grad}\to0.
\label{eq:GRcriterion_main_DchiDgrad}
\end{equation}
In this limit one has
\begin{equation}
\chi_{\FF}\to0,
\quad
q_a^{(g)}\to0,
\quad
\pi_{ab}^{(g)}\to0,
\quad
\chi_A\,\D_a^{(\FF)}\phi^A\to0.
\label{eq:GRcriterion_consequences}
\end{equation}

Thus, when $\mathcal{B}_{AB}$ is nondegenerate and positive definite, the thermal system may be described in terms of the set $(\chi_{\FF},W_a,\mathfrak D_\chi,\mathfrak D_{\rm grad})$. Accordingly, Eq.~\eqref{eq:Dchi_transport_sigma} should be interpreted as the dynamical equation governing the time-like part of the GR approach, while $\mathfrak D_{\rm grad}$ captures the complementary spatial part. The multi-scalar attractor problem is therefore not exhausted by the coupling variable $\chi_{\FF}$: it is the coupled disappearance of the full field-space thermal vector and of the residual scalar spatial structure that characterizes relaxation to the GR sector.

\section{Entropy current and entropy production in the coupling frame}\label{sec:entropy}

Having identified the natural thermal variables of the coupling channel and their transport system, one may now ask how the corresponding entropy current is modified by the presence of the remaining scalar directions. A fully generic-frame entropy analysis would remain partly formal unless the compatibility condition for the existence of a local temperature is satisfied. For this reason, we restrict attention here to the $\FF$-comoving frame, where the variables $\chi_{\FF}$ and $W_a^{(\FF)}$ are already defined and the associated transport system is explicit. As in the one-field literature, one must additionally choose a normalization for the pair $(K_{\FF},T_{\FF})$ once their product $\chi_{\FF}=K_{\FF}T_{\FF}$ has been fixed. With that understood, we follow the same first-order Eckart construction adopted in Ref.~\cite{Faraoni:2021jri} and define the entropy current of the geometric effective fluid by
\begin{equation}
s^a = s\,u^a + \frac{q^a}{T},
\label{eq:entropycurrentGHMP}
\end{equation}
where $s$ is the entropy density and $T$ is the effective temperature. Assuming a closed (though not isolated) effective system, the entropy density is obtained from the first law in the form
\begin{equation}
s=\frac{\rho_g+p_g}{T}.
\label{eq:entropydensityGHMP}
\end{equation}

Using Eq.~\eqref{eq:qFfull}, the entropy-generation vector is
\begin{equation}
R_a^{(\FF)}
\equiv
\frac{q_a^{(g)}}{T_{\FF}}
=
-K_{\FF}\left(a_a^{(\FF)}+W_a^{(\FF)}\right).
\label{eq:entropygenerationvectorGHMP}
\end{equation}
This already exhibits the main new thermodynamic feature of the generic multi-scalar case. In the one-field theory the entropy-generating part of the current is purely inertial, being controlled only by the acceleration of the scalar-comoving congruence. Here, by contrast, it contains the additional contribution $W_a^{(\FF)}$, which measures the residual temperature-gradient sector generated by the scalar directions not eliminated by the coupling-frame choice. Thus the entropy current is no longer determined by the inertial channel alone: it is sensitive to the remaining field-space thermal structure.

Using the coupling-frame expressions for $\rho_g$ and $p_g$, one finds
\begin{align}
(\rho_g+p_g)_{(\FF)}
&=
\frac{1}{8\pi \FF}
\left[
\ddot\FF_{(\FF)}
-\frac13\Theta_{(\FF)}\dot\FF_{(\FF)}
\right.
\nonumber\\
&\left.
+\mathcal{B}_{AB}\dot{\phi}^A_{(\FF)}\dot{\phi}^B_{(\FF)}
+\frac13\mathcal{B}_{AB}\D_a^{(\FF)}\phi^A\D^a_{(\FF)}\phi^B
\right].
\label{eq:rhoppluspFF}
\end{align}
Hence the entropy density of the geometric multi-scalar medium is
\begin{align}
s_{(\FF)}
&=
\frac{K_{\FF}}{\chi_{\FF}}(\rho_g+p_g)_{(\FF)}
\nonumber\\
&=
-\frac{K_{\FF}}{\dot\FF_{(\FF)}}
\left[
\ddot\FF_{(\FF)}
-\frac13\Theta_{(\FF)}\dot\FF_{(\FF)}
\right.
\nonumber\\
&\left.
+\mathcal{B}_{AB}\dot{\phi}^A_{(\FF)}\dot{\phi}^B_{(\FF)}
+\frac13\mathcal{B}_{AB}\D_a^{(\FF)}\phi^A\D^a_{(\FF)}\phi^B
\right].
\label{eq:entropydensityFFexplicit}
\end{align}
or in terms of time-like thermal magnitude and spatial gradient magnitude quantities
\begin{align}
s_{(\FF)}&=-\frac{K_{\FF}}{\dot\FF_{(\FF)}}\left[\ddot\FF_{(\FF)}-\frac13\Theta_{(\FF)}\dot\FF_{(\FF)}+64\pi^2\FF^2\,\mathfrak D_\chi\right.\nonumber\\
&\left.+\frac13\,\mathfrak D_{\rm grad}
\right].
\label{eq:entropydensityFF_DchiDgrad}
\end{align}

Thus, even in the coupling frame, the entropy density is not controlled solely by the coupling channel: it receives explicit contributions from the kinetic and spatial-gradient sectors of the remaining scalar degrees of freedom.

If the effective particle current is conserved, $\nabla_a(nu^a)=0$, the first-order entropy production law gives~\cite{Eckart:1940te,Faraoni:2021jri}
\begin{equation}
\nabla_a s^a
=
\frac{\Pi_g^{\,2}}{\zeta_{\FF}T_{\FF}}
+
\frac{q_a^{(g)}q^a_{(g)}}{K_{\FF}T_{\FF}^{\,2}}
+
\frac{\pi_{ab}^{(g)}\pi_{(g)}^{ab}}{2\eta_{\FF}T_{\FF}}.
\label{eq:entropyproductionGHMPgeneral}
\end{equation}

The expression above is the first-order entropy-production functional of the matched effective geometric medium. It is not a proof that every effective-fluid rewriting of the multi-scalar field equations produces nonnegative entropy. Nonnegative entropy production requires the usual first-order thermodynamic sign assumptions. In particular, one must restrict to branches with positive effective temperature,
\begin{equation}
T_{\FF}>0,
\end{equation} 
nonnegative heat conductivity,
\begin{equation}
K_{\FF}\ge0,
\end{equation}
and, whenever the corresponding sectors are retained, nonnegative effective bulk and shear viscosities,
\begin{equation}
\zeta_{\FF}\ge0,
\qquad
\eta_{\FF}\ge0.
\end{equation}
If these sign conditions fail, the expression remains a formally defined entropy-production functional of the matched effective medium, but it should not be interpreted as satisfying a second-law statement.

Using Eq.~\eqref{eq:qFfull}, the heat-flux contribution becomes
\begin{equation}
\frac{q_a^{(g)}q^a_{(g)}}{K_{\FF}T_{\FF}^{\,2}}
=
K_{\FF}\left(a_a^{(\FF)}+W_a^{(\FF)}\right)
\left(a^a_{(\FF)}+W^a_{(\FF)}\right),
\label{eq:qtermentropyGHMP}
\end{equation}
and therefore the entropy production of the multi-scalar medium is
\begin{align}
\nabla_a s^a
&=
\frac{\Pi_g^{\,2}}{\zeta_{\FF}T_{\FF}}
+
K_{\FF}\left(a_a^{(\FF)}+W_a^{(\FF)}\right)
\left(a^a_{(\FF)}+W^a_{(\FF)}\right)
\nonumber\\
&
+
\frac{\pi_{ab}^{(g)}\pi_{(g)}^{ab}}{2\eta_{\FF}T_{\FF}}.
\label{eq:entropyproductionGHMPexpanded}
\end{align}

This is the central entropy-production formula of the generic multi-scalar theory. Relative to the one-field case, the new effect is not merely the presence of more scalar fields, but the appearance of a genuinely additional nonequilibrium sector. Expanding the middle term shows that the entropy production contains three distinct contributions: a purely inertial piece $K_{\FF}a_a a^a$, a genuinely multi-scalar temperature-gradient piece $K_{\FF}W_aW^a$, and the mixed term $2K_{\FF}a_aW^a$, which couples the inertial and temperature-gradient channels. Hence the additional scalar directions contribute to entropy production not only by supplying extra energy density and pressure, but also by generating an additional irreversible thermal structure.

If one adopts the economical convention $\Pi_g=0$, Eq.~\eqref{eq:entropyproductionGHMPexpanded} reduces to
\begin{equation}
\nabla_a s^a
=
K_{\FF}\left(a_a^{(\FF)}+W_a^{(\FF)}\right)
\left(a^a_{(\FF)}+W^a_{(\FF)}\right)
+
\frac{\pi_{ab}^{(g)}\pi_{(g)}^{ab}}{2\eta_{\FF}T_{\FF}}.
\label{eq:entropyproductionGHMPnobulk}
\end{equation}
Using the shear constitutive relation in the $\FF$-frame,
\begin{equation}
\pi_{ab}^{(g)}=-2\eta_{\FF}\,\sigma_{ab}^{(\FF)},
\label{eq:piEckartEntropyFF}
\end{equation}
one obtains
\begin{equation}
\frac{\pi_{ab}^{(g)}\pi_{(g)}^{ab}}{2\eta_{\FF}T_{\FF}}
=
\frac{2\eta_{\FF}}{T_{\FF}}\,
\sigma_{ab}^{(\FF)}\sigma^{ab}_{(\FF)}
=
K_{\FF}\,\frac{2\eta_{\FF}}{\chi_{\FF}}\,
\sigma_{ab}^{(\FF)}\sigma^{ab}_{(\FF)}.
\label{eq:sheartermentropyFF2}
\end{equation}
If, moreover, in the case where the matching condition ~\eqref{eq:piCondF_multi} is satisfied then Eq.~\eqref{eq:entropyproductionGHMPnobulk} becomes
\begin{align}
\nabla_a s^a
&=
K_{\FF}\Bigg[
\left(a_a^{(\FF)}+W_a^{(\FF)}\right)\left(a^a_{(\FF)}+W^a_{(\FF)}\right)
\nonumber\\
&
+\left(\frac{\Lambda_{(\FF)}}{\dot\FF_{(\FF)}}-1\right)
\sigma_{ab}^{(\FF)}\sigma^{ab}_{(\FF)}
\Bigg].
\label{eq:entropyproductionGHMPnobulkSigma}
\end{align}
Accordingly, the inertial/temperature-gradient contribution is manifestly nonnegative, whereas the sign of the shear contribution is controlled by the effective viscosity $\eta_{\FF}$.

If instead one adopts the minimal bulk split, Eq.~\eqref{eq:PiFF} and Eq.~\eqref{eq:zetaFF}, then
\begin{align}
\nabla_a s^a
&=
K_{\FF}\Bigg[
\frac{2}{3}\Theta_{(\FF)}^{2}
+\left(a_a^{(\FF)}+W_a^{(\FF)}\right)\left(a^a_{(\FF)}+W^a_{(\FF)}\right)
\nonumber\\
&
+\left(\frac{\Lambda_{(\FF)}}{\dot\FF_{(\FF)}}-1\right)
\sigma_{ab}^{(\FF)}\sigma^{ab}_{(\FF)}
\Bigg].
\label{eq:entropyproductionGHMPminbulkSigma2}
\end{align}

In this interpretation the entropy budget is naturally organized into three channels: a positive bulk contribution, a positive inertial/temperature-gradient contribution, and a shear contribution whose sign depends on the effective viscosity.

On the purely inertial branch $W_a^{(\FF)}=0$, the entropy-generation vector reduces to
\begin{equation}
R_a^{(\FF)}=-K_{\FF}a_a^{(\FF)},
\label{eq:entropygenerationvectorGHMPinertial}
\end{equation}
and the entropy production becomes
\begin{equation}
\nabla_a s^a
=
\frac{\Pi_g^{\,2}}{\zeta_{\FF}T_{\FF}}
+
K_{\FF}a_a^{(\FF)}a^a_{(\FF)}
+
\frac{\pi_{ab}^{(g)}\pi_{(g)}^{ab}}{2\eta_{\FF}T_{\FF}}.
\label{eq:entropyproductionGHMPInertial}
\end{equation}
This provides the expected consistency check: once the residual temperature-gradient sector disappears, the entropy current and entropy production collapse to the familiar one-field structure.

Furthermore, in the aligned-gradient sector one has $W_a^{(\FF)}=0$ and
\begin{equation}
\eta_{\FF}=-\frac{\chi_{\FF}}{2},
\end{equation}
so that
\begin{equation}
\frac{\pi_{ab}^{(g)}\pi_{(g)}^{ab}}{2\eta_{\FF}T_{\FF}}
=
-K_{\FF}\sigma_{ab}\sigma^{ab}.
\label{eq:sheartermentropyaligned}
\end{equation}
With the economical bulk choice this yields
\begin{equation}
\nabla_a s^a
=
K_{\FF}\left(a_a a^a-\sigma_{ab}\sigma^{ab}\right).
\label{eq:entropyproductionaligned}
\end{equation}
This is the direct multi-scalar analogue of the one-field result, now recovered on the special branch where all scalar gradients align and the multichannel system effectively collapses to a single scalar-comoving thermal medium.

The entropy analysis therefore makes the role of the additional scalar directions especially transparent. At the level of the current, they add a new contribution through $W_a^{(\FF)}$. At the level of the density, they contribute directly through the time-like thermal magnitude and spatial gradient magnitude in Eq.~\eqref{eq:entropydensityFF_DchiDgrad}. At the level of entropy production, they generate a genuinely new irreversible channel through the $W_a^{(\FF)}$-dependent part of Eq.~\eqref{eq:entropyproductionGHMPexpanded}. Thus, even at first order, the generic multi-scalar theory is not merely a one-field thermal system with modified coefficients: it possesses an enlarged nonequilibrium sector tied to the residual field-space dynamics that survives after the coupling channel has been put in inertial form.

\section{Homogeneous and isotropic cosmology: coupling channel and multi-scalar diagnostics}
\label{sec:cosmoKTF}

A natural application of the formalism developed until now is a spatially homogeneous and isotropic universe. In the present context, the cosmological sector is especially useful because it isolates the coupling thermal variable $\chi_{\FF}$ in the cleanest possible setting while also showing, in a particularly transparent way, that the full multi-scalar thermal dynamics is not exhausted by this single quantity. Indeed, Friedmann--Lema\^itre--Robertson--Walker (FLRW) symmetry suppresses the residual temperature-gradient sector $W_a^{(\FF)}$ identically and removes the entire spatial scalar structure, but it does \emph{not} eliminate the field-space thermal vector $\chi^A$. Thus homogeneous cosmology realizes the simplest branch of the multi-scalar transport system at the level of the heat flux, without collapsing the underlying scalar dynamics to the one-field case. This also makes clear the limitation of the FLRW application. Exact homogeneity and isotropy remove the projected spatial gradients by symmetry. Therefore FLRW tests the coupling channel and the time-like multi-scalar thermal vector, but it does not activate the genuinely spatial sector encoded in $W_a^{(\FF)}$, $\mathfrak D_{\rm grad}$, and the projected-gradient transport equation. This limitation is structural rather than technical: a less symmetric geometry is needed for the spatial sector to be present. Natural next applications are Bianchi cosmologies, where anisotropic expansion can activate shear and direction-dependent scalar gradients, and Lema\^itre--Tolman--Bondi geometries, where radial inhomogeneity can source nontrivial projected scalar gradients. These cases provide natural settings in which to test the full multi-scalar spatial sector beyond FLRW.

Consider a FLRW line element
\begin{equation}
\mathrm{d}s^2=-\mathrm{d}t^2+a^2(t)\,\gamma_{ij}\mathrm{d}x^i \mathrm{d}x^j,
\label{eq:FLRWmetric}
\end{equation}
with comoving cosmological flow $u^a=\delta^a{}_t$ and Hubble parameter
\begin{equation}
H\equiv \frac{\dot a}{a}.
\end{equation}
Spatial homogeneity implies
\begin{equation}
\D_a\phi^A=0,
\qquad
\D_a\FF=0,
\label{eq:homogeneousscalars}
\end{equation}
while isotropy gives
\begin{equation}
a_a=0,
\qquad
\omega_{ab}=0,
\qquad
\sigma_{ab}=0,
\qquad
\Theta=3H.
\label{eq:FLRWkinematics}
\end{equation}
Hence the geometric medium is necessarily of perfect-fluid form,
\begin{equation}
q_a^{(g)}=0,
\qquad
\pi_{ab}^{(g)}=0.
\label{eq:FLRWqpi}
\end{equation}

The thermodynamic interpretation of these statements is immediate. First, from Eq.~\eqref{eq:WF_multi},
\begin{equation}
W_a^{(\FF)}
=
-\frac{1}{\dot\FF}\,
\mathcal{B}_{AB}\dot\phi^A\,\D_a\phi^B
=0,
\label{eq:WFLRWzero}
\end{equation}
so the explicit coupling-frame heat-flux system $(\chi_{\FF},W_a^{(\FF)})$ collapses in FLRW to the single scalar variable $\chi_{\FF}$. Second, the spatial diagnostic $\mathfrak D_{\rm grad}$ vanishes identically,
\begin{equation}
\mathfrak D_{\rm grad}^{\rm FLRW}=0,
\qquad
\dot{\mathfrak D}_{\rm grad}^{\rm FLRW}=0,
\label{eq:DgradFLRWzero}
\end{equation}
because the projected scalar gradients vanish by symmetry. Thus the entire spatial part of the GR-attractor problem is removed kinematically in exact FLRW. What remains is the time-like multi-scalar sector.

Using the general expressions for $\rho_g$ and $p_g$, the geometric effective-fluid variables reduce to
\begin{align}
\rho_g^{\mathrm{FLRW}}
&=
\frac{1}{8\pi \FF}
\left[
-3H\dot\FF
+\frac12\,\mathcal{B}_{AB}\dot{\phi}^A\dot{\phi}^B
+\frac{U}{2}
\right],
\label{eq:rhoFLRW}
\\
p_g^{\mathrm{FLRW}}
&=
\frac{1}{8\pi \FF}
\left[
\ddot\FF+2H\dot\FF
+\frac12\,\mathcal{B}_{AB}\dot{\phi}^A\dot{\phi}^B
-\frac{U}{2}
\right].
\label{eq:pFLRW}
\end{align}
Thus, even though the background heat flux and anisotropic stress vanish identically, the scalar sector still contributes nontrivially through the kinetic invariant $\mathcal{B}_{AB}\dot\phi^A\dot\phi^B$.

It is useful to rewrite these expressions in terms of the field-space thermal variables $\chi_A$ and $\chi^A$ yielding
\begin{align}
\rho_g^{\mathrm{FLRW}}
&=
3H\chi_{\FF}
+4\pi\FF\,\chi_A\chi^A
+\frac{U}{16\pi\FF},
\label{eq:rhoFLRW_chiA}
\\
p_g^{\mathrm{FLRW}}
&=
\frac{\ddot\FF+2H\dot\FF}{8\pi\FF}
+4\pi\FF\,\chi_A\chi^A
-\frac{U}{16\pi\FF}.
\label{eq:pFLRW_chiA}
\end{align}
When $\mathcal{B}_{AB}$ is nondegenerate, the combination $\chi_A\chi^A$ is precisely the time-like diagnostic $\mathfrak D_\chi$ so that the FLRW effective fluid depends on the multi-scalar state through two conceptually distinct quantities: the coupling projection $\chi_{\FF}$ and the full time-like invariant $\mathfrak D_\chi$. If, in addition, $\mathcal{B}_{AB}$ is positive definite, then $\mathfrak D_\chi$ is nonnegative and has the interpretation of a genuine norm-like measure of the magnitude of the full scalar velocity vector in field space.

We assume that the matter sector is described by a perfect fluid,
\begin{equation}
T_{ab}^{(m)}=(\rho_m+p_m)u_a u_b+p_m g_{ab},
\label{eq:TmFLRW}
\end{equation}
for which
\begin{equation}
T^{(m)}=-\rho_m+3p_m.
\label{eq:TmtraceFLRW}
\end{equation}
Matter conservation then gives
\begin{equation}
\dot\rho_m+3H(\rho_m+p_m)=0.
\label{eq:matterconsFLRW}
\end{equation}

If $\nabla_a\FF$ is timelike and aligned with the cosmological flow, then the FLRW congruence coincides with the $\FF$-comoving frame so that, in the homogeneous sector,
\begin{equation}
\chi_{\FF}=0
\quad\Longleftrightarrow\quad
\dot\FF=0
\quad\Longleftrightarrow\quad
\nabla_a\FF=0.
\label{eq:chiFequivFLRW}
\end{equation}
This exact equivalence is special to FLRW, because $\D_a\FF=0$ identically; it should not be confused with the more general inhomogeneous case, where $\chi_{\FF}\to0$ is only a near-equilibrium statement. Eq.~\eqref{eq:chiFtransportSigma} reduces to the cosmological transport law
\begin{equation}
\dot\chi_{\FF}
=
8\pi\chi_{\FF}^{\,2}
-3H\chi_{\FF}
+\Sigma_{\FF}^{\mathrm{FLRW}},
\label{eq:KTFtransportFLRW}
\end{equation}
where
\begin{equation}
\Sigma_{\FF}^{\mathrm{FLRW}}
\equiv
\left.\widetilde\Sigma_{\FF}\right|_{\mathrm{FLRW}}.
\label{eq:SigmaFFLRW}
\end{equation}

If one wishes to write the source explicitly in terms of field-space quantities, one may introduce
\begin{equation}
Q\equiv \mathcal B^{AB}\FF_{,A}\FF_{,B},
\
P\equiv \mathcal B^{AB}\FF_{,A}U_{,B},
\
\mathcal H_{AB}\equiv \nabla_A\nabla_B\FF,
\end{equation}
so that
\begin{widetext}
\begin{equation}
\Sigma_{\FF}^{\mathrm{FLRW}}
=
\frac{
-8\pi\FF\,\mathcal H_{AB}\chi^A\chi^B
+\dfrac{Q}{2\FF}\left(\dfrac{T^{(m)}}{\FF}+8\pi\FF\,\mathfrak D_\chi-\dfrac{U}{4\pi\FF}\right)
+\dfrac{P}{16\pi\FF}
}{
1+\dfrac{3Q}{2\FF}
}.
\label{eq:SigmaFFLRWexplicit_chiA}
\end{equation}
\end{widetext}
This form makes the multi-scalar structure explicit: even in exact homogeneity, the coupling thermal variable $\chi_{\FF}$ is sourced not only by the Hessian of the coupling and by matter, but also by the full time-like invariant $\mathfrak D_\chi$.

The evolution of the full time-like multi-scalar state is governed by the FLRW reduction of Eq.~\eqref{eq:Dchi_transport_sigma}. Since $\D_a\phi^A=0$ identically, all spatial-gradient contributions disappear and one obtains
\begin{equation}
\dot{\mathfrak D}_\chi
=
2\left(8\pi\chi_{\FF}-3H\right)\mathfrak D_\chi
+\Sigma_{\chi}^{\mathrm{FLRW}},
\label{eq:Dchi_transport_FLRW}
\end{equation}
with
\begin{equation}
\Sigma_{\chi}^{\mathrm{FLRW}}
\equiv
\frac{1}{8\pi\FF}\,\chi_A\left(U^{,A}-\FF^{,A}R\right).
\label{eq:SigmaChiFLRW}
\end{equation}
Accordingly, the homogeneous cosmological dynamics retains the full distinction between the coupling channel and the complete time-like multi-scalar state. The variable $\chi_{\FF}$ still controls the coupling projection, whereas $\mathfrak D_\chi$ determines whether the magnitude of the full scalar velocity vector in field space is shrinking or growing.

A particularly transparent subcase is the simultaneous source-free branch
\begin{equation}
\Sigma_{\FF}^{\mathrm{FLRW}}=0,
\qquad
\Sigma_{\chi}^{\mathrm{FLRW}}=0.
\label{eq:FLRW_sourcefree_pair}
\end{equation}
Then
\begin{equation}
\dot\chi_{\FF}
=
\chi_{\FF}\left(8\pi\chi_{\FF}-3H\right),
\label{eq:KTFsimpleFLRW}
\end{equation}
and
\begin{equation}
\dot{\mathfrak D}_\chi
=
2\mathfrak D_\chi\left(8\pi\chi_{\FF}-3H\right).
\label{eq:Dchi_simple_FLRW}
\end{equation}
Thus, for an expanding universe, $H>0$, one has
\begin{equation}
0<\chi_{\FF}<\frac{3H}{8\pi}
\quad\Longrightarrow\quad
\dot\chi_{\FF}<0,
\qquad
\dot{\mathfrak D}_\chi<0,
\label{eq:FLRW_double_cooling}
\end{equation}
so both the coupling channel and the full time-like multi-scalar state are damped, whereas
\begin{equation}
\chi_{\FF}>\frac{3H}{8\pi}
\quad\Longrightarrow\quad
\dot\chi_{\FF}>0,
\qquad
\dot{\mathfrak D}_\chi>0.
\label{eq:FLRW_double_heating}
\end{equation}
For a contracting universe, $H<0$, every positive $\chi_{\FF}$ gives
\begin{equation}
\dot\chi_{\FF}>0,
\qquad
\dot{\mathfrak D}_\chi>0,
\label{eq:FLRW_contracting_growth}
\end{equation}
so contraction universally amplifies both the coupling channel and the full time-like multi-scalar state on this branch. These relations are the exact homogeneous counterparts of the general transport analysis developed in Sec.~\ref{sec:transport}. 

Finally, in FLRW the entropy current is purely convective,
\begin{equation}
s^a_{\rm FLRW}=s_{\rm FLRW}\,u^a,
\label{eq:entropycurrentFLRW}
\end{equation}
with
\begin{equation}
s_{\rm FLRW}
=
-\frac{K_{\FF}}{\dot\FF}
\left[
\ddot\FF-H\dot\FF+\mathcal{B}_{AB}\dot\phi^A\dot\phi^B
\right],
\label{eq:entropydensityFLRW}
\end{equation}
or, equivalently,
\begin{equation}
s_{\rm FLRW}
=
-\frac{K_{\FF}}{\dot\FF}
\left[
\ddot\FF-H\dot\FF+64\pi^2\FF^2\,\mathfrak D_\chi
\right].
\label{eq:entropydensityFLRW_chiA}
\end{equation}
Thus, even though all spatial dissipative fluxes vanish by symmetry and $\mathfrak D_{\rm grad}$ is identically zero, the entropy density still depends explicitly on the full time-like multi-scalar state through $\mathfrak D_\chi$.

To summarize, FLRW cosmology suppresses the residual temperature-gradient sector and removes the spatial attractor problem altogether, but it does not trivialize the full multi-scalar thermal dynamics. The explicit coupling-frame heat-flux description collapses to the single variable $\chi_{\FF}$ because $W_a^{(\FF)}=0$, yet the background still carries the nontrivial field-space thermal invariant $\mathfrak D_\chi$. The cosmological sector therefore provides a particularly clean realization of the central multi-scalar lesson: the one-field thermal picture survives only at the level of the coupling projection, whereas the full time-like scalar thermal state remains intrinsically multidimensional.

\section{Discussion and conclusions}
\label{sec:discussion}

In this work we developed a first-order thermodynamic description of general Jordan-frame tensor--multi-scalar gravity, keeping the common coupling function $\FF(\phi^A)$ and the field-space kinetic matrix $\mathcal{B}_{AB}(\phi^C)$ explicit throughout. Our aim was to determine how the recent thermodynamic interpretation of one-field scalar-tensor gravity extends to the genuinely multi-field regime. The main conclusion is that the multi-field problem is not a trivial index extension of the one-field case. Once the scalar sector is promoted to a nontrivial field space, different scalar directions can play distinct dynamical and constitutive roles, and the thermodynamic description must therefore distinguish between the coupling channel, the full time-like scalar sector, and the residual spatial scalar structure.

At the structural level, the Einstein-like form of the field equations, Eq.~\eqref{eq:einsteinlike_multi}, together with the exact covariant $1+3$ decomposition of the geometric source, shows that the scalar sector behaves as an effective imperfect medium. The corresponding variables, Eqs.~\eqref{eq:rhog_multi}--\eqref{eq:pig_multi_multi}, split naturally into coupling, kinetic, and potential contributions. In particular, the heat flux, Eq.~\eqref{eq:qg_multi}, and the anisotropic stress, Eq.~\eqref{eq:pig_multi_multi}, are not controlled by the coupling function alone, but also contain genuinely multi-field contributions weighted by $\mathcal{B}_{AB}$. Thus, even before any constitutive interpretation is imposed, tensor--multi-scalar gravity already exhibits a multichannel effective-fluid structure with no strict one-field analogue.

A central result of the paper is the generic-frame constitutive analysis. Matching the theory-defined heat flux to Eckart's law, Eqs.~\eqref{eq:eckart}--\eqref{eq:eckartp}, shows that the existence of an effective temperature is not automatic, but requires a nontrivial integrability condition, Eq.~\eqref{eq:compatibility_app}, or Eq.~\eqref{eq:compatibility_irrot_app} in the irrotational case. In a multi-field theory this obstruction is especially significant, because no single scalar-comoving frame generically removes all scalar-gradient contributions simultaneously. The effective heat sector therefore need not reduce to a purely inertial channel. In the frame adapted to the effective coupling $\FF$, this becomes particularly transparent: the heat flux takes the form~\eqref{eq:heatFF}, with the coupling variable $\chi_{\FF}=K_{\FF}T_{\FF}$ defined in Eq.~\eqref{eq:XiF_multi} and the additional spatial contribution encoded in the covector $W_a^{(\FF)}$, Eq.~\eqref{eq:WF_multi}. Thus the natural thermal description is generically not a $KT$-only theory, but a broader constitutive system in which the coupling channel and the residual temperature-gradient structure coexist.

This leads to one of the main conclusions of the present work: in tensor--multi-scalar gravity, relaxation of the coupling channel is not equivalent to full relaxation toward the GR sector. The variable $\chi_{\FF}$ probes only the component of the scalar dynamics along the coupling direction selected by $\FF_{,A}$, cf.\ Eqs.~\eqref{eq:chiFraw} and \eqref{eq:chiFfromchiA_transport}, and its transport equation~\eqref{eq:chiFtransportSigma} therefore diagnoses the heating or cooling of that channel alone. The full time-like multi-scalar state is instead encoded in the field-space thermal vector $\chi^A$, Eq.~\eqref{eq:chiAdef_transport}, and in its constitutive covector $\chi_A$, Eq.~\eqref{eq:chiAlowerdef_transport}. When the kinetic matrix is nondegenerate, the scalar $\mathfrak D_\chi$, Eq.~\eqref{eq:Dchi_def}, provides the canonical invariant contraction of this sector; if $\mathcal{B}_{AB}$ is moreover positive definite, Eq.~\eqref{eq:Dchi_positive}, it becomes a genuine norm-like measure whose evolution is governed by Eq.~\eqref{eq:Dchi_transport_sigma}. Even this, however, is not sufficient to characterize approach to the GR sector, because the residual spatial scalar structure must also decay. This is captured by the second diagnostic $\mathfrak D_{\rm grad}$, Eq.~\eqref{eq:Dgrad_def}, whose evolution is governed by Eq.~\eqref{eq:Dgrad_transport_sigma} and which measures the projected scalar-gradient sector when $\mathcal{B}_{AB}$ is positive definite. The GR-like endpoint is therefore characterized not by the coupling condition alone, but by the joint disappearance of the time-like and spatial sectors, as summarized by Eqs.~\eqref{eq:GRcriterion_main}--\eqref{eq:GRcriterion_main_DchiDgrad}. In this sense, the multi-field extension of the thermal attractor-to-GR picture is genuinely richer than in the one-field case: freezing the effective coupling is only a partial notion of relaxation. Although the formalism can be formulated in greater generality, its strongest and most transparent thermodynamic realization is obtained when the field-space kinetic matrix $\mathcal{B}_{AB}$ is nondegenerate and positive definite, since only in that case do the diagnostics $\mathfrak D_\chi$ and $\mathfrak D_{\rm grad}$ acquire a canonical and genuinely positive-definite interpretation as measures of the time-like and spatial multi-scalar sectors. 

The entropy analysis reinforces this picture. In the coupling frame, the entropy current~\eqref{eq:entropycurrentGHMP} and the entropy-generation vector~\eqref{eq:entropygenerationvectorGHMP} are not determined solely by the inertial channel. Through $W_a^{(\FF)}$, they are also sensitive to the residual thermal structure that survives after adapting the frame to the coupling. Likewise, the entropy density, Eqs.~\eqref{eq:entropydensityFFexplicit} and \eqref{eq:entropydensityFF_DchiDgrad}, depends explicitly on both the time-like thermal magnitude $\mathfrak D_\chi$ and the spatial gradient magnitude $\mathfrak D_{\rm grad}$. At the level of entropy production, Eqs.~\eqref{eq:entropyproductionGHMPexpanded}, \eqref{eq:entropyproductionGHMPnobulkSigma}, and \eqref{eq:entropyproductionGHMPminbulkSigma2}, the multi-field theory therefore possesses an enlarged nonequilibrium sector relative to the one-field case. The gravitational medium is thermodynamically richer not merely because there are more scalar degrees of freedom, but because these degrees of freedom can contribute through distinct constitutive channels that need not collapse to a single effective variable.

A further qualification concerns the kinetic matrix $\mathcal B_{AB}$. The formalism distinguishes between a general symmetric kinetic matrix, a nondegenerate kinetic matrix, and a positive-definite kinetic matrix. The lower-index scalar equations and the effective stress tensor remain meaningful for a general symmetric $\mathcal B_{AB}$. The field-space covariant rewriting of the scalar equations and the raising of field-space indices require nondegeneracy. The interpretation of $\mathfrak D_\chi$ and $\mathfrak D_{\rm grad}$ as nonnegative norm-like diagnostics requires the stronger assumption of positive definiteness. If $\mathcal B_{AB}$ is degenerate, the theory may still be studied, but the analysis must either be kept in lower-index form or supplemented by additional model-dependent structure.

The FLRW cosmology example provides a particularly transparent illustration of the general picture. By symmetry, $W_a^{(\FF)}$ and $\mathfrak D_{\rm grad}$ vanish identically, Eqs.~\eqref{eq:WFLRWzero} and \eqref{eq:DgradFLRWzero}. However, the theory does not reduce fully to the one-field thermal case, because the time-like invariant $\mathfrak D_\chi$ remains nontrivial, enters the effective fluid variables through Eqs.~\eqref{eq:rhoFLRW_chiA} and \eqref{eq:pFLRW_chiA}, and contributes to the coupling transport law through Eq.~\eqref{eq:SigmaFFLRWexplicit_chiA}. The homogeneous dynamics is therefore still governed by both Eqs.~\eqref{eq:KTFtransportFLRW} and \eqref{eq:Dchi_transport_FLRW}, showing that coupling equilibrium, Eq.~\eqref{eq:chiFequivFLRW}, does not in general imply full multi-scalar equilibrium. Moreover, the FLRW example is therefore limited. It is useful because it shows that the homogeneous theory does not reduce completely to the one-field thermal case: the full time-like field-space invariant $\mathfrak D_\chi$ remains active. However, FLRW removes the spatial multi-scalar sector by symmetry, since $W_a^{(\FF)}=0$ and $\mathfrak D_{\rm grad}=0$. A natural next step is to apply the present framework to less symmetric systems. Bianchi cosmologies are especially useful because anisotropic expansion makes the shear sector active and can couple nontrivially to the multi-scalar spatial gradients. Lema\^itre--Tolman--Bondi geometries are another natural arena, because radial inhomogeneity provides a simple setting in which the projected scalar gradients need not vanish. These examples would allow one to test the spatial transport sector of the formalism, which is suppressed by FLRW symmetry.

Taken together, these results show that tensor--multi-scalar gravity should be viewed thermodynamically as a genuinely multichannel gravitational medium. What survives from the recent one-field program is the privileged role of the coupling channel and of its associated inertial variable $\chi_{\FF}$. What is genuinely new is that this channel no longer exhausts the thermal state of the theory. In the multi-field case, the thermodynamic interpretation of scalar-tensor gravity must distinguish between coupling-channel heating and cooling, controlled by Eq.~\eqref{eq:chiFtransportSigma}, full time-like multi-scalar relaxation, controlled by Eq.~\eqref{eq:Dchi_transport_sigma}, and the evolution of the residual spatial scalar structure, controlled by Eq.~\eqref{eq:Dgrad_transport_sigma}. This is, in our view, the main physical message of the paper.

From a broader perspective, the thermal interpretation of modified gravity becomes more discriminating in the multi-field regime: a configuration may look close to GR at the level of the effective coupling while still carrying hidden nonequilibrium scalar structure in field-space directions orthogonal to $\FF_{,A}$. This distinction is potentially relevant for cosmological attractors, for the interpretation of screening or relaxation mechanisms, and for model building in situations where several scalar degrees of freedom remain active even when the effective gravitational coupling is nearly frozen. The formalism developed here is designed precisely to isolate these possibilities in a covariant and thermodynamically transparent way. In this sense, it provides not merely a reinterpretation of the field equations, but a diagnostic framework able to distinguish apparent GR-like behavior at the level of the effective coupling from genuine relaxation of the full multi-scalar sector. 

Because the analysis was carried out for a general Jordan-frame tensor-multi-scalar theory, the formalism applies well beyond any single model and provides a common thermodynamic language for a broad class of multi-field scalar-tensor constructions. This makes the framework useful not only for formal analysis, but also for comparing different multi-field models on a common constitutive and transport footing. In particular, it makes clear which thermodynamic features are genuinely multi-field and which arise only in special aligned sectors that effectively collapse to the one-field case. More broadly, the framework developed here opens the way to several natural extensions: applications to specific multi-field models and exact solutions; the analysis of stronger-field and compact-object configurations; the development of a second-order causal completion of the coupled thermal system; and the investigation of whether the multi-scalar relaxation criteria identified here can be related systematically to observationally relevant attractor behavior in cosmology and astrophysics.

\begin{acknowledgments}
We thank the anonymous referee for the valuable suggestions provided, which have greatly contributed to enhancing the quality of this manuscript. We thank Jess Rutschi, Jos\'{e} Pedro Mimoso, Francisco Lobo, Leonor Ferro and Miguel Pinto for helpful comments. This work was supported by Fundação para a Ciência e a Tecnologia (FCT) through national funds under the research grant UID/04434/2025 (DOI 10.54499/UID/04434/2025).
\end{acknowledgments}

\bibliographystyle{apsrev4-2}
\bibliography{bibliog}

@article{Will:2014kxa,
  author       = {Will, Clifford M.},
  title        = {The Confrontation between General Relativity and Experiment},
  journal      = {Living Rev. Rel.},
  volume       = {17},
  pages        = {4},
  year         = {2014},
  doi          = {10.12942/lrr-2014-4},
  eprint       = {1403.7377},
  archivePrefix= {arXiv},
  primaryClass = {gr-qc}
}

@article{Berti:2015itd,
  author       = {Berti, Emanuele and Barausse, Enrico and Cardoso, Vitor and Gualtieri, Leonardo and Pani, Paolo and Sperhake, Ulrich and Stein, Leo C. and Wex, Norbert and Yagi, Kent and Baker, Tessa and others},
  title        = {Testing General Relativity with Present and Future Astrophysical Observations},
  journal      = {Class. Quant. Grav.},
  volume       = {32},
  pages        = {243001},
  year         = {2015},
  doi          = {10.1088/0264-9381/32/24/243001},
  eprint       = {1501.07274},
  archivePrefix= {arXiv},
  primaryClass = {gr-qc}
}

@article{Clifton:2011jh,
  author       = {Clifton, Timothy and Ferreira, Pedro G. and Padilla, Antonio and Skordis, Constantinos},
  title        = {Modified Gravity and Cosmology},
  journal      = {Phys. Rept.},
  volume       = {513},
  pages        = {1--189},
  year         = {2012},
  doi          = {10.1016/j.physrep.2012.01.001},
  eprint       = {1106.2476},
  archivePrefix= {arXiv},
  primaryClass = {astro-ph.CO}
}

@book{CANTATA:2021asi,
  author       = {Akrami, Yashar and others},
  editor       = {Saridakis, Emmanuel N. and Lazkoz, Ruth and Salzano, Vincenzo and Vargas Moniz, Paulo and Capozziello, Salvatore and Beltr{\'a}n Jim{\'e}nez, Jose and De Laurentis, Mariafelicia and Olmo, Gonzalo J.},
  collaboration = {CANTATA},
  title        = {Modified Gravity and Cosmology. An Update by the CANTATA Network},
  publisher    = {Springer},
  year         = {2021},
  doi          = {10.1007/978-3-030-83715-0},
  eprint       = {2105.12582},
  archivePrefix= {arXiv},
  primaryClass = {gr-qc}
}

@article{CosmoVerseNetwork:2025alb,
  author       = {Di Valentino, Eleonora and others},
  collaboration = {CosmoVerse Network},
  title        = {The CosmoVerse White Paper: Addressing observational tensions in cosmology with systematics and fundamental physics},
  journal      = {Phys. Dark Univ.},
  volume       = {49},
  pages        = {101965},
  year         = {2025},
  doi          = {10.1016/j.dark.2025.101965},
  eprint       = {2504.01669},
  archivePrefix= {arXiv},
  primaryClass = {astro-ph.CO}
}

@article{Brans:1961sx,
  author       = {Brans, C. and Dicke, R. H.},
  title        = {Mach's principle and a relativistic theory of gravitation},
  journal      = {Phys. Rev.},
  volume       = {124},
  pages        = {925--935},
  year         = {1961},
  doi          = {10.1103/PhysRev.124.925}
}

@article{Horndeski:1974wa,
  author       = {Horndeski, Gregory Walter},
  title        = {Second-order scalar-tensor field equations in a four-dimensional space},
  journal      = {Int. J. Theor. Phys.},
  volume       = {10},
  pages        = {363--384},
  year         = {1974},
  doi          = {10.1007/BF01807638}
}

@book{Faraoni:2004pi,
  author       = {Faraoni, Valerio},
  title        = {Cosmology in Scalar-Tensor Gravity},
  year         = {2004},
  doi          = {10.1007/978-1-4020-1989-0}
}

@article{Damour:1992we,
  author       = {Damour, Thibault and Esposito-Farese, Gilles},
  title        = {Tensor-multi-scalar theories of gravitation},
  journal      = {Class. Quant. Grav.},
  volume       = {9},
  pages        = {2093--2176},
  year         = {1992},
  doi          = {10.1088/0264-9381/9/9/015}
}

@incollection{Wands:2007bd,
  author       = {Wands, David},
  title        = {Multiple field inflation},
  booktitle    = {The Cosmic Microwave Background and Physics of the Early Universe},
  series       = {Lect. Notes Phys.},
  volume       = {738},
  pages        = {275--304},
  year         = {2008},
  doi          = {10.1007/978-3-540-74353-8_8},
  eprint       = {astro-ph/0702187},
  archivePrefix= {arXiv}
}

@article{Byrnes:2006fr,
  author       = {Byrnes, Christian T. and Wands, David},
  title        = {Curvature and isocurvature perturbations from two-field inflation in a slow-roll expansion},
  journal      = {Phys. Rev. D},
  volume       = {74},
  pages        = {043529},
  year         = {2006},
  doi          = {10.1103/PhysRevD.74.043529},
  eprint       = {astro-ph/0605679},
  archivePrefix= {arXiv}
}

@article{Rosa:2021lhc,
  author       = {Rosa, Jo{\~a}o Lu{\'\i}s and Lobo, Francisco S. N. and Olmo, Gonzalo J.},
  title        = {Weak-field regime of the generalized hybrid metric-Palatini gravity},
  journal      = {Phys. Rev. D},
  volume       = {104},
  number       = {12},
  pages        = {124030},
  year         = {2021},
  doi          = {10.1103/PhysRevD.104.124030},
  eprint       = {2104.10890},
  archivePrefix= {arXiv},
  primaryClass = {gr-qc}
}

@article{Rosa:2017jld,
  author       = {Rosa, Jo{\~a}o Lu{\'\i}s and Carloni, Sante and Lemos, Jos{\'e} Pizarro de Sande e and Lobo, Francisco Sab{\'e}lio Nobrega},
  title        = {Cosmological solutions in generalized hybrid metric-Palatini gravity},
  journal      = {Phys. Rev. D},
  volume       = {95},
  number       = {12},
  pages        = {124035},
  year         = {2017},
  doi          = {10.1103/PhysRevD.95.124035},
  eprint       = {1703.03335},
  archivePrefix= {arXiv},
  primaryClass = {gr-qc}
}

@article{Rosa:2019ejh,
  author       = {Rosa, Jo{\~a}o L. and Carloni, Sante and Lemos, Jos{\'e} P. S.},
  title        = {Cosmological phase space of generalized hybrid metric-Palatini theories of gravity},
  journal      = {Phys. Rev. D},
  volume       = {101},
  number       = {10},
  pages        = {104056},
  year         = {2020},
  doi          = {10.1103/PhysRevD.101.104056},
  eprint       = {1908.07778},
  archivePrefix= {arXiv},
  primaryClass = {gr-qc}
}

@article{Rosa:2021mln,
  author       = {Rosa, Jo{\~a}o Lu{\'\i}s and Lemos, Jos{\'e} P. S.},
  title        = {Junction conditions for generalized hybrid metric-Palatini gravity with applications},
  journal      = {Phys. Rev. D},
  volume       = {104},
  number       = {12},
  pages        = {124076},
  year         = {2021},
  doi          = {10.1103/PhysRevD.104.124076},
  eprint       = {2111.12109},
  archivePrefix= {arXiv},
  primaryClass = {gr-qc}
}

@article{Rosa:2018jwp,
  author       = {Rosa, Jo{\~a}o Lu{\'\i}s and Lemos, Jos{\'e} P. S. and Lobo, Francisco S. N.},
  title        = {Wormholes in generalized hybrid metric-Palatini gravity obeying the matter null energy condition everywhere},
  journal      = {Phys. Rev. D},
  volume       = {98},
  number       = {6},
  pages        = {064054},
  year         = {2018},
  doi          = {10.1103/PhysRevD.98.064054},
  eprint       = {1808.08975},
  archivePrefix= {arXiv},
  primaryClass = {gr-qc}
}

@article{Aliannejadi:2024byu,
  author       = {Aliannejadi, Reyhaneh and Haghani, Zahra},
  title        = {Quark Stars in Generalized Hybrid Metric-Palatini Gravity},
  journal      = {Iran. J. Astron. Astrophys.},
  volume       = {11},
  number       = {1},
  pages        = {45--57},
  year         = {2024},
  doi          = {10.22128/ijaa.2024.819.1183},
  eprint       = {2410.00159},
  archivePrefix= {arXiv},
  primaryClass = {gr-qc}
}

@article{Gomes:2025ers,
  author       = {Gomes, Cl{\'a}udio and Rosa, Jo{\~a}o Lu{\'\i}s and Pinto, Miguel A. S.},
  title        = {Gravitational wave propagation in generalized hybrid metric-Palatini gravity},
  journal      = {Eur. Phys. J. C},
  volume       = {85},
  number       = {11},
  pages        = {1359},
  year         = {2025},
  doi          = {10.1140/epjc/s10052-025-15085-x},
  eprint       = {2506.12870},
  archivePrefix= {arXiv},
  primaryClass = {gr-qc}
}

@article{Faraoni:2018qdr,
  author       = {Faraoni, Valerio and Cot{\'e}, Jeremy},
  title        = {Imperfect fluid description of modified gravities},
  journal      = {Phys. Rev. D},
  volume       = {98},
  number       = {8},
  pages        = {084019},
  year         = {2018},
  doi          = {10.1103/PhysRevD.98.084019},
  eprint       = {1808.02427},
  archivePrefix= {arXiv},
  primaryClass = {gr-qc}
}

@article{Faraoni:2021lfc,
    author = "Faraoni, Valerio and Giusti, Andrea",
    title = "{Thermodynamics of scalar-tensor gravity}",
    eprint = "2103.05389",
    archivePrefix = "arXiv",
    primaryClass = "gr-qc",
    doi = "10.1103/PhysRevD.103.L121501",
    journal = "Phys. Rev. D",
    volume = "103",
    number = "12",
    pages = "L121501",
    year = "2021"
}

@article{Faraoni:2021jri,
  author       = {Faraoni, Valerio and Giusti, Andrea and Mentrelli, Andrea},
  title        = {New approach to the thermodynamics of scalar-tensor gravity},
  journal      = {Phys. Rev. D},
  volume       = {104},
  number       = {12},
  pages        = {124031},
  year         = {2021},
  doi          = {10.1103/PhysRevD.104.124031},
  eprint       = {2110.02368},
  archivePrefix= {arXiv},
  primaryClass = {gr-qc}
}

@article{FaraoniGiustiEtAl2022Peculiar,
  author       = {Faraoni, Valerio and Giusti, Andrea and Jose, Sonia and Giardino, Serena},
  title        = {Peculiar thermal states in the first-order thermodynamics of gravity},
  journal      = {Phys. Rev. D},
  volume       = {106},
  pages        = {024049},
  year         = {2022},
  doi          = {10.1103/PhysRevD.106.024049},
  eprint       = {2206.02046},
  archivePrefix= {arXiv},
  primaryClass = {gr-qc}
}

@article{Faraoni:2022gry,
  author       = {Faraoni, Valerio and Giardino, Serena and Giusti, Andrea and Vanderwee, Robert},
  title        = {Scalar field as a perfect fluid: thermodynamics of minimally coupled scalars and Einstein frame scalar-tensor gravity},
  journal      = {Eur. Phys. J. C},
  volume       = {83},
  number       = {1},
  pages        = {24},
  year         = {2023},
  doi          = {10.1140/epjc/s10052-023-11186-7},
  eprint       = {2208.04051},
  archivePrefix= {arXiv},
  primaryClass = {gr-qc}
}

@article{Giusti:2021sku,
    author = "Giusti, Andrea and Zentarra, Stefan and Heisenberg, Lavinia and Faraoni, Valerio",
    title = "{First-order thermodynamics of Horndeski gravity}",
    eprint = "2108.10706",
    archivePrefix = "arXiv",
    primaryClass = "gr-qc",
    doi = "10.1103/PhysRevD.105.124011",
    journal = "Phys. Rev. D",
    volume = "105",
    number = "12",
    pages = "124011",
    year = "2022"
}

@article{Faraoni:2023hwu,
  author       = {Faraoni, Valerio and Houle, Julien},
  title        = {More on the first-order thermodynamics of scalar-tensor and Horndeski gravity},
  journal      = {Eur. Phys. J. C},
  volume       = {83},
  number       = {6},
  pages        = {521},
  year         = {2023},
  doi          = {10.1140/epjc/s10052-023-11712-7},
  eprint       = {2302.01442},
  archivePrefix= {arXiv},
  primaryClass = {gr-qc}
}

@article{Giardino:2023ygc,
  author       = {Giardino, Serena and Giusti, Andrea},
  title        = {First-order thermodynamics of scalar-tensor gravity},
  journal      = {Ric. Mat.},
  volume       = {74},
  number       = {1},
  pages        = {43--59},
  year         = {2025},
  doi          = {10.1007/s11587-023-00801-0},
  eprint       = {2306.01580},
  archivePrefix= {arXiv},
  primaryClass = {gr-qc}
}

@article{Miranda:2024dhw,
  author       = {Miranda, Marcello and Giardino, Serena and Giusti, Andrea and Heisenberg, Lavinia},
  title        = {First-order thermodynamics of Horndeski cosmology},
  journal      = {Phys. Rev. D},
  volume       = {109},
  number       = {12},
  pages        = {124033},
  year         = {2024},
  doi          = {10.1103/PhysRevD.109.124033},
  eprint       = {2401.10351},
  archivePrefix= {arXiv},
  primaryClass = {gr-qc}
}

@article{Houle:2024sxs,
  author       = {Houle, Julien and Faraoni, Valerio},
  title        = {New phenomenology in the first-order thermodynamics of scalar-tensor gravity for Bianchi universes},
  journal      = {Phys. Rev. D},
  volume       = {110},
  number       = {2},
  pages        = {024067},
  year         = {2024},
  doi          = {10.1103/PhysRevD.110.024067},
  eprint       = {2404.19470},
  archivePrefix= {arXiv},
  primaryClass = {gr-qc}
}

@article{Faraoni:2025alq,
  author       = {Faraoni, Valerio and Giusti, Andrea},
  title        = {Thermal Origin of the Attractor-to-General-Relativity in Scalar-Tensor Gravity},
  journal      = {Phys. Rev. Lett.},
  volume       = {134},
  number       = {21},
  pages        = {211406},
  year         = {2025},
  doi          = {10.1103/22w4-v2xn},
  eprint       = {2502.18272},
  archivePrefix= {arXiv},
  primaryClass = {gr-qc}
}

@article{Faraoni:2025ufi,
  author       = {Faraoni, Valerio},
  title        = {Black hole interiors in the thermal view of scalar-tensor gravity},
  journal      = {Phys. Rev. D},
  volume       = {112},
  number       = {2},
  pages        = {L021504},
  year         = {2025},
  doi          = {10.1103/mk89-hjkn},
  eprint       = {2505.08322},
  archivePrefix= {arXiv},
  primaryClass = {gr-qc}
}

@article{Faraoni:2025dex,
  author       = {Faraoni, Valerio and Veilleux, Nikki},
  title        = {Thermal view of singularity-free scalar-tensor spacetimes},
  journal      = {Phys. Rev. D},
  volume       = {113},
  number       = {4},
  pages        = {044030},
  year         = {2026},
  doi          = {10.1103/bf3m-8ydv},
  eprint       = {2511.04941},
  archivePrefix= {arXiv},
  primaryClass = {gr-qc}
}

@article{Faraoni:2025fjq,
  author       = {Faraoni, Valerio and Cattivelli, Santiago Novoa},
  title        = {Thermal view of f(R) cosmology},
  journal      = {Phys. Rev. D},
  volume       = {113},
  number       = {4},
  pages        = {044034},
  year         = {2026},
  doi          = {10.1103/y86x-18t8},
  eprint       = {2511.00347},
  archivePrefix= {arXiv},
  primaryClass = {gr-qc}
}

@article{Bhattacharyya:2025tgp,
  author       = {Bhattacharyya, Soham and SenGupta, Soumitra},
  title        = {Thermal description of braneworld effective theories},
  journal      = {Phys. Rev. D},
  volume       = {113},
  number       = {6},
  pages        = {064019},
  year         = {2026},
  doi          = {10.1103/dyf3-h6gb},
  eprint       = {2508.14228},
  archivePrefix= {arXiv},
  primaryClass = {hep-th}
}

@article{Miranda:2022uyk,
  author       = {Miranda, Marcello and Graham, Pierre-Antoine and Faraoni, Valerio},
  title        = {Effective fluid mixture of tensor-multi-scalar gravity},
  journal      = {Eur. Phys. J. Plus},
  volume       = {138},
  number       = {5},
  pages        = {387},
  year         = {2023},
  doi          = {10.1140/epjp/s13360-023-03984-5},
  eprint       = {2211.03958},
  archivePrefix= {arXiv},
  primaryClass = {gr-qc}
}

@article{Eckart:1940te,
  author       = {Eckart, C.},
  title        = {The Thermodynamics of irreversible processes. 3. Relativistic theory of the simple fluid},
  journal      = {Phys. Rev.},
  volume       = {58},
  pages        = {919--924},
  year         = {1940},
  doi          = {10.1103/PhysRev.58.919}
}

@article{Hiscock:1985zz,
  author       = {Hiscock, William A. and Lindblom, Lee},
  title        = {Generic instabilities in first-order dissipative relativistic fluid theories},
  journal      = {Phys. Rev. D},
  volume       = {31},
  pages        = {725--733},
  year         = {1985},
  doi          = {10.1103/PhysRevD.31.725}
}

@inproceedings{Maartens:1996vi,
  author       = {Maartens, Roy},
  title        = {Causal thermodynamics in relativity},
  eprint       = {astro-ph/9609119},
  archivePrefix= {arXiv},
  year         = {1996}
}

@article{Grana:2005jc,
  author       = {Gra\~na, Mariana},
  title        = {Flux compactifications in string theory: A comprehensive review},
  journal      = {Phys. Rept.},
  volume       = {423},
  pages        = {91--158},
  year         = {2006},
  doi          = {10.1016/j.physrep.2005.10.008},
  eprint       = {hep-th/0509003},
  archivePrefix= {arXiv}
}

@article{Pereira:2025dmk,
    author = "Pereira, David S. and Mimoso, Jos{\'e} Pedro",
    title = "{Eckart heat-flux applicability in $F(Φ,X)R$ theories and the existence of temperature gradients}",
    eprint = "2512.20553",
    archivePrefix = "arXiv",
    primaryClass = "gr-qc",
    month = "12",
    journal="",
    year = "2025"
}

@article{Pereira:2025rsr,
    author = "Pereira, David S. and Capozziello, Salvatore and Lobo, Francisco S. N. and Mimoso, Jos{\'e} Pedro",
    title = "{Novel scalar degrees of freedom emerging from hybrid metric-Palatini gravity}",
    eprint = "2511.09208",
    archivePrefix = "arXiv",
    primaryClass = "gr-qc",
    doi = "10.1103/pzxl-dm4q",
    journal = "Phys. Rev. D",
    volume = "113",
    number = "6",
    pages = "064055",
    year = "2026"
}

@article{Starobinsky:2001xq,
  author       = {Starobinsky, Alexei A. and Tsujikawa, Shinji and Yokoyama, Jun'ichi},
  title        = {Cosmological perturbations from multi-field inflation in generalized Einstein theories},
  journal      = {Nucl. Phys. B},
  volume       = {610},
  pages        = {383--410},
  year         = {2001},
  doi          = {10.1016/S0550-3213(01)00322-4},
  eprint       = {astro-ph/0107555},
  archivePrefix= {arXiv},
  primaryClass = {astro-ph}
}

@article{Giardino:2022sdv,
    author = "Giardino, Serena and Faraoni, Valerio and Giusti, Andrea",
    title = "{First-order thermodynamics of scalar-tensor cosmology}",
    eprint = "2202.07393",
    archivePrefix = "arXiv",
    primaryClass = "gr-qc",
    doi = "10.1088/1475-7516/2022/04/053",
    journal = "JCAP",
    volume = "04",
    number = "04",
    pages = "053",
    year = "2022"
}

@article{Kaiser:2012ak,
  author       = {Kaiser, David I. and Mazenc, Edward A. and Sfakianakis, Evangelos I.},
  title        = {Primordial Bispectrum from Multifield Inflation with Nonminimal Couplings},
  journal      = {Phys. Rev. D},
  volume       = {87},
  pages        = {064004},
  year         = {2013},
  doi          = {10.1103/PhysRevD.87.064004},
  eprint       = {1210.7487},
  archivePrefix= {arXiv},
  primaryClass = {astro-ph.CO}
}

@article{Kaiser:2013sna,
  author       = {Kaiser, David I. and Sfakianakis, Evangelos I.},
  title        = {Multifield Inflation after Planck: The Case for Nonminimal Couplings},
  journal      = {Phys. Rev. Lett.},
  volume       = {112},
  pages        = {011302},
  year         = {2014},
  doi          = {10.1103/PhysRevLett.112.011302},
  eprint       = {1304.0363},
  archivePrefix= {arXiv},
  primaryClass = {astro-ph.CO}
}

@article{Karamitsos:2017elm,
  author       = {Karamitsos, Sotirios and Pilaftsis, Apostolos},
  title        = {Frame Covariant Nonminimal Multifield Inflation},
  journal      = {Nucl. Phys. B},
  volume       = {927},
  pages        = {219--254},
  year         = {2018},
  doi          = {10.1016/j.nuclphysb.2017.12.015},
  eprint       = {1706.07011},
  archivePrefix= {arXiv},
  primaryClass = {hep-ph}
}

@article{DeCross:2015uza,
  author       = {DeCross, Matthew P. and Kaiser, David I. and Prabhu, Anirudh and Prescod-Weinstein, Chanda and Sfakianakis, Evangelos I.},
  title        = {Preheating after Multifield Inflation with Nonminimal Couplings. I. Covariant Formalism and Attractor Behavior},
  journal      = {Phys. Rev. D},
  volume       = {97},
  pages        = {023526},
  year         = {2018},
  doi          = {10.1103/PhysRevD.97.023526},
  eprint       = {1510.08553},
  archivePrefix= {arXiv},
  primaryClass = {astro-ph.CO}
}

@article{DeCross:2016fdz,
  author       = {DeCross, Matthew P. and Kaiser, David I. and Prabhu, Anirudh and Prescod-Weinstein, Chanda and Sfakianakis, Evangelos I.},
  title        = {Preheating after Multifield Inflation with Nonminimal Couplings. II. Resonance Structure},
  journal      = {Phys. Rev. D},
  volume       = {97},
  pages        = {023527},
  year         = {2018},
  doi          = {10.1103/PhysRevD.97.023527},
  eprint       = {1610.08868},
  archivePrefix= {arXiv},
  primaryClass = {astro-ph.CO}
}

@article{DeCross:2016xpj,
  author       = {DeCross, Matthew P. and Kaiser, David I. and Prabhu, Anirudh and Prescod-Weinstein, Chanda and Sfakianakis, Evangelos I.},
  title        = {Preheating after Multifield Inflation with Nonminimal Couplings. III. Dynamical Spacetime Results},
  journal      = {Phys. Rev. D},
  volume       = {97},
  pages        = {023528},
  year         = {2018},
  doi          = {10.1103/PhysRevD.97.023528},
  eprint       = {1610.08916},
  archivePrefix= {arXiv},
  primaryClass = {astro-ph.CO}
}

@article{Nguyen:2019kbm,
  author       = {Nguyen, Rachel and van de Vis, Jorinde and Sfakianakis, Evangelos I. and Giblin, John T. and Kaiser, David I.},
  title        = {Nonlinear Dynamics of Preheating after Multifield Inflation with Nonminimal Couplings},
  journal      = {Phys. Rev. Lett.},
  volume       = {123},
  pages        = {171301},
  year         = {2019},
  doi          = {10.1103/PhysRevLett.123.171301},
  eprint       = {1905.12562},
  archivePrefix= {arXiv},
  primaryClass = {astro-ph.CO}
}

@article{Christodoulidis:2025wew,
    author = "Christodoulidis, Perseas and Rosati, Robert and Sfakianakis, Evangelos I.",
    title = "{Robust non-minimal attractors in many-field inflation}",
    eprint = "2504.12406",
    archivePrefix = "arXiv",
    primaryClass = "astro-ph.CO",
    month = "4",
    journal="",
    year = "2025"
}

@article{Bolshakova:2025xay,
    author = "Bolshakova, Katerina and Chervon, Sergey",
    title = "{A New Approach to the Analysis of Cosmological Parameters in Multifield Cosmology}",
    eprint = "2501.01948",
    archivePrefix = "arXiv",
    primaryClass = "gr-qc",
    reportNumber = "LGCA-USPU 12/2024",
    month = "1",
    journal="",
    year = "2025"
}

@article{Brax:2026ttj,
    author = "Brax, Philippe and van de Bruck, Carsten and Davis, Anne-Christine and Smith, Adam",
    title = "{Multi-Field Dilaton Screening Beyond the Thin-Shell Mechanism}",
    eprint = "2603.13986",
    archivePrefix = "arXiv",
    primaryClass = "gr-qc",
    month = "3",
    journal="",
    year = "2026"
}

@article{Akrami:2020zlw,
  author       = {Akrami, Yashar and Sasaki, Misao and Solomon, Adam R. and Vardanyan, Valeri},
  title        = {Multi-field dark energy: cosmic acceleration on a steep potential},
  journal      = {Phys. Lett. B},
  volume       = {819},
  pages        = {136427},
  year         = {2021},
  doi          = {10.1016/j.physletb.2021.136427},
  eprint       = {2008.13660},
  archivePrefix= {arXiv},
  primaryClass = {astro-ph.CO}
}

@article{Smith:2024mbp,
  author       = {Smith, Adam and Mylova, Maria and Brax, Philippe and van de Bruck, Carsten and Burgess, C. P. and Davis, Anne-Christine},
  title        = {CMB Implications of Multi-field Axio-dilaton Cosmology},
  journal      = {JCAP},
  volume       = {12},
  pages        = {058},
  year         = {2024},
  doi          = {10.1088/1475-7516/2024/12/058},
  eprint       = {2408.10820},
  archivePrefix= {arXiv},
  primaryClass = {hep-th}
}

@article{Smith:2024mna,
  author       = {Smith, Adam and Mylova, Maria and Brax, Philippe and van de Bruck, Carsten and Burgess, C. P. and Davis, Anne-Christine},
  title        = {A Minimal Axio-dilaton Dark Sector},
  journal      = {JCAP},
  volume       = {07},
  pages        = {023},
  year         = {2025},
  doi          = {10.1088/1475-7516/2025/07/023},
  eprint       = {2410.11099},
  archivePrefix= {arXiv},
  primaryClass = {hep-th}
}

@article{Smith:2025wxd,
  author       = {Smith, Adam and Brax, Philippe and van de Bruck, Carsten and Burgess, C. P. and Davis, Anne-Christine},
  title        = {Screened Axio-dilaton Cosmology: Novel Forms of Early Dark Energy},
  journal      = {Eur. Phys. J. C},
  volume       = {85},
  pages        = {1008},
  year         = {2025},
  doi          = {10.1140/epjc/s10052-025-14735-4},
  eprint       = {2505.05450},
  archivePrefix= {arXiv},
  primaryClass = {hep-th}
}

@article{Gallego:2026xqq,
    author = "Gallego, Diego and Orjuela-Quintana, J. Bayron",
    title = "{Multifield dark energy: Interplay between curved field space and curved spacetime}",
    eprint = "2603.18341",
    archivePrefix = "arXiv",
    primaryClass = "hep-th",
    month = "3",
    journal="",
    year = "2026"
}

@article{Hohmann:2016yfd,
    author = "Hohmann, Manuel and Jarv, Laur and Kuusk, Piret and Randla, Erik and Vilson, Ott",
    title = "{Post-Newtonian parameter $\gamma$ for multiscalar-tensor gravity with a general potential}",
    eprint = "1607.02356",
    archivePrefix = "arXiv",
    primaryClass = "gr-qc",
    doi = "10.1103/PhysRevD.94.124015",
    journal = "Phys. Rev. D",
    volume = "94",
    number = "12",
    pages = "124015",
    year = "2016"
}

@article{Berkin:1993bt,
    author = "Berkin, Andrew L. and Hellings, Ronald W.",
    title = "{Multiple field scalar - tensor theories of gravity and cosmology}",
    eprint = "gr-qc/9401033",
    archivePrefix = "arXiv",
    reportNumber = "WU-AP-10-91",
    doi = "10.1103/PhysRevD.49.6442",
    journal = "Phys. Rev. D",
    volume = "49",
    pages = "6442--6449",
    year = "1994"
}

@article{Tsagas:2007yx,
    author = "Tsagas, Christos G. and Challinor, Anthony and Maartens, Roy",
    title = "{Relativistic cosmology and large-scale structure}",
    eprint = "0705.4397",
    archivePrefix = "arXiv",
    primaryClass = "astro-ph",
    doi = "10.1016/j.physrep.2008.03.003",
    journal = "Phys. Rept.",
    volume = "465",
    pages = "61--147",
    year = "2008"
}

@article{Ellis:1971pg,
    author = "Ellis, G. F. R.",
    title = "{Relativistic cosmology}",
    doi = "10.1007/s10714-009-0760-7",
    journal = "Proc. Int. Sch. Phys. Fermi",
    volume = "47",
    pages = "104--182",
    year = "1971"
}

@article{Schon:2021uhz,
  author       = {Sch{\"o}n, Oliver and Doneva, Daniela D.},
  title        = {Tensor-Multi-Scalar Gravity: Equations of Motion to 2.5 post-Newtonian Order},
  journal      = {Phys. Rev. D},
  volume       = {105},
  pages        = {064034},
  year         = {2022},
  doi          = {10.1103/PhysRevD.105.064034},
  eprint       = {2112.07388},
  archivePrefix= {arXiv},
  primaryClass = {gr-qc}
}

@article{Jarv:2026dbb,
    author = {J{\"a}rv, Laur and Karamitsos, Sotirios},
    title = "{Frame invariant diffusive formulation of scalar-tensor gravity}",
    journal="",
    eprint = "2604.16094",
    archivePrefix = "arXiv",
    primaryClass = "gr-qc",
    month = "4",
    year = "2026"
}

@article{Banerjee:2026ulx,
    author = "Banerjee, Narayan and Giusti, Andrea and Faraoni, Valerio and Gallerani, Luca and Maltais-Gosselin, Alain",
    title = "{Towards a causal effective thermodynamics of scalar-tensor gravity}",
    eprint = "2605.24133",
    archivePrefix = "arXiv",
    primaryClass = "gr-qc",
    journal="", 
    month = "5",
    year = "2026"
}

\end{document}